\documentclass[a4paper,12pt]{article}
\pdfoutput=1 

\usepackage{jheppub} 

\usepackage[T1]{fontenc} 
\usepackage{bm,latexsym,amsmath,amssymb,amsfonts,mathtools,mathrsfs}
\usepackage{comment}
\usepackage{ulem}

\usepackage{color}
\usepackage{hyperref}

\usepackage{todonotes}

\usepackage[framemethod=TikZ]{mdframed}
\usepackage{lipsum}

\mdfdefinestyle{MyFrame}{
linecolor=blue,
outerlinewidth=2pt,
roundcorner=20pt,
innertopmargin=\baselineskip,
innerbottommargin=\baselineskip,
innerrightmargin=20pt,
innerleftmargin=20pt,
backgroundcolor=gray!20!white}

\newcommand{\be}{\begin{equation}}
\newcommand{\ee}{\end{equation}}
\newcommand{\beal}{\begin{aligned}}
\newcommand{\eeal}{\end{aligned}}
\newcommand\bea {\begin{eqnarray}}
\newcommand\eea {\end{eqnarray}}
\newcommand{\bec}{\begin{cases}}
\newcommand{\eec}{\end{cases}}

\title{Black Holes, Oscillating Instantons, and the Hawking-Moss transition}

\author[a,b,c]{Ruth Gregory,}
\author[d]{Ian G. Moss}
\author[c]{and Naritaka Oshita}

\emailAdd{r.a.w.gregory@durham.ac.uk}
\emailAdd{ian.moss@newcastle.ac.uk}
\emailAdd{noshita@pitp.ca}

\affiliation[a]{Institute for Particle Physics Phenomenology, Department of Physics, 
Durham University, South Road, Durham DH1 3LE, UK}
\affiliation[b]{Department of Mathematical Sciences, Durham University,
South Road, Durham, DH1 3LE, UK}
\affiliation[c]{Perimeter Institute, 31 Caroline Street North, Waterloo, 
ON, N2L 2Y5, Canada}
\affiliation[d]{School of Mathematics, Statistics and Physics, Newcastle University, 
Newcastle Upon Tyne, NE1 7RU, UK}

\abstract{
Static oscillating bounces in Schwarzschild de Sitter spacetime are investigated. 
The oscillating bounce with many oscillations gives a super-thick bubble wall, 
for which the total vacuum energy increases while the mass of the black hole 
decreases due to the conservation of Arnowitt-Deser-Misner (ADM) mass. 
We show that the transition rate of such an ``up-tunneling" consuming the seed 
black hole is higher than that of the Hawking-Moss transition. The 
correspondence of analyses in the static and global coordinates in the 
Euclidean de Sitter space is also investigated.
}

\begin{document}
\maketitle
\flushbottom

\section{Introduction}

Cosmological phase transitions involving supercooling and the formation 
of bubbles may have played an important role in the early universe. 
In the case of extreme supercooling, the phase transition involves a 
quantum transition of a scalar field through a potential barrier from a 
false vacuum state to a true vacuum state. The vacuum decay rate is
conventionally described in terms of an instanton, or bounce solution, 
to the field equations in imaginary time \cite{coleman1977,callan1977,CDL}.
In an interesting recent twist, primordial black holes \cite{Gregory:2013hja,
Burda:2015isa,Burda:2015yfa,Burda:2016mou,Mukaida:2017bgd,Oshita:2019jan} 
and horizonless compact objects \cite{Oshita:2018ptr,Koga:2019mee} 
have been shown to act as nucleation seeds for true vacuum bubbles, and 
significantly enhance the decay rate (see also early work 
\cite{Hiscock:1987hn,Berezin:1987ea,Berezin:1990qs}).
There is however another type of tunnelling solution -- the Hawking-Moss (HM)
instanton \cite{Hawking:1981fz} -- that is relevant for relatively flat
potentials, and the aim of this paper is to explore how seeded nucleation
proceeds in this case.

We are interested in situations where gravitational effects on bubble 
nucleation are important. This was first investigated by Coleman and 
de Luccia (CDL) \cite{CDL}, who looked at bounce solutions
to the Einstein-scalar system with $O(4)$ symmetry. Shortly after their work, 
in the context of the inflationary universe, Hawking and Moss noticed that
CDL bounce solutions did not exist for very flat potentials, and instead 
suggested an $O(5)$ symmetric bounce solution, the HM instanton, 
was the appropriate solution. The nucleation rate obtained this way 
has since been supported by quantum cosmology \cite{Hartle:1983ai}
and by stochastic methods \cite{Starobinsky:1986fx}. The HM instanton 
is static, and consists of a solution purely at the top of the potential barrier. 
These features are shared by bounce solutions in thermal systems, and a 
modern interpretation of the HM instanton is that it represents a thermal 
fluctuation within a horizon volume at the temperature of de Sitter (dS)
space \cite{Brown:2007sd}.

The condition for the existence of CDL instantons can be given explicitly in 
terms of the scalar potential $V$ and its second derivative at the top of the 
potential barrier \cite{Hawking:1981fz,Hackworth:2004xb,
Balek:2004sd,Weinberg:2005af,Kanno:2011vm,Battarra:2013rba}
\begin{equation}
\left|{d^2V\over d\phi^2}\right|> 4 \left( { 8 \pi V\over 3 M_{\rm Pl}^2} \right) 
\label{bound}
\end{equation}
where $M_{\rm Pl} \equiv 1/ \sqrt{G}$ is the Planck mass.
This condition is relaxed when the fourth derivative $d^4V/d\phi^4$ is subdominant \cite{Demetrian:2005ag}\footnote{In addition, non-minimal Higgs-gravity 
interaction may lead to an ambiguity in the definition of potential curvature 
$V'' (\phi_{\rm top})$ \cite{Rajantie:2016hkj,Stopyra:2018cjy,Joti:2017fwe}, 
which might change the CDL bound (\ref{bound}).}.
Furthermore, close to this bound there are typically multiple solutions 
of an oscillatory type which oscillate back and forth between either side 
of the potential barrier \cite{Hackworth:2004xb,Lavrelashvili:2006cv,Lee:2012qv,
Lee:2014ula}. Like the HM instanton,
these may be regarded as thermal excitations, however, it has been 
suggested that the oscillatory solutions have more than one negative mode 
and because of this cannot represent true vacuum decay \cite{Lee:2014ula}.

Seeded vacuum decay by a primordial black hole \cite{Burda:2015isa,
Burda:2015yfa,Burda:2016mou} is another case which has a close 
resemblance to thermally assisted quantum tunnelling, however this time 
at the Hawking black hole temperature. In fact, the tunnelling rate can be 
expressed thermodynamically using Boltzmann's formula 
\cite{Gregory:2013hja,Oshita:2016oqn},
\begin{equation}
\Gamma=Ae^{\Delta {\cal S}},
\label{boltzmann}
\end{equation}
where $A$ is the pre-factor and $\Delta {\cal S}$ is the change in 
black hole (BH) entropy due to the loss 
of mass during the tunnelling event. (The entropy can go down in a fluctuation, 
which is why this is an unlikely event). The BHs break some of the 
symmetry, so that the bounce solutions have $O(3)$ symmetry.
However, the bounces that dominate the decay rate are also 
imaginary-time translation invariant, as we would expect from the thermal context.
These bounces were investigated primarily in the context of Higgs decay in an 
asymptotically flat universe, however, here we examine the more general question
of when seeded bounce solutions exist, and present new results on seeded HM
tunnelling. We are therefore looking at the early universe in situations
where the potential barrier is close to the limit in Eq.\ (\ref{bound})\footnote{A 
\textit{local} symmetry restoration due to the Hawking radiation, that exhibits 
an up transition around a BH, was discussed in \cite{Moss:1984zf,Flachi:2011sx,
Oshita:2016btk} where thermal backreaction to the effective potential assists 
the up transition. On the other hand, in this manuscript, we are interested in 
a \textit{global} up tunneling inside the cosmological horizon, such as the HM 
transition, without strong thermal backreaction to effective potential.}.

In this manuscript, we investigate oscillating bounces in the Schwarzschild 
de Sitter (SdS) background using static coordinates. The organization 
of this manuscript is as follows: Before embarking on the analysis in the static 
SdS background, in Sec.\ \ref{sec_analytic_bounce} we first investigate analytically
the oscillating bounce solutions, both for a dS background as well as close to the
Nariai limit \cite{Nariai:1,Nariai:2}, using the static SdS co-ordinates. This
contrasts to the conventional analysis, such as in \cite{Hackworth:2004xb},
that are performed in \textit{global} dS coordinates. We carefully show that 
the analysis in both static and global dS patches lead to the same eigenvalue 
restrictions on $V'' (\phi_{\rm top})/(M_{\rm Pl}^2 V)$ (where a prime denotes 
the derivative with respect to $\phi$) for a probe solution, as well as the same 
functional form for the scalar perturbation. This is a highly nontrivial result,
as the analytic continuation of the global and static patches does not
coincide, and of course the coverage of the static patch in the Lorentzian section
is not the full dS spacetime. Sec.\ \ref{sec_SdS_numerical} presents the oscillating 
bounce solutions in a general SdS background, obtained by solving the (non-linear)
Einstein-scalar field equations with time-translation and spherical symmetry. 
Then we constrain the parameter region of the mass of a seed BH and 
$V'' (\phi_{\rm top})/(M_{\rm Pl}^2 V)$ in which the oscillating bounces exist. 
In Sec.\ \ref{sec_HM} we discuss the HM bounce around a BH, that conserves 
the total energy inside the cosmological horizon, and compare its bounce 
configuration and Euclidean action with those of the oscillating bounce 
with many oscillations. Then we conclude that the higher-mode oscillating 
bounces may be regarded as the intermediate bounces between 
the fundamental bounce (no turning point for $\phi$) and the HM bounce 
around a BH. We conclude in the final section.

\section{Oscillating Bounce in Static Patch: dS and Nariai Limits}
\label{sec_analytic_bounce}

The $O(4)$ symmetric oscillating bounce solutions were analysed close 
to the critical limit of \eqref{bound} by Hackworth and Weinberg 
\cite{Hackworth:2004xb}. The argument runs as follows: Any bounce 
solution interpolating between vacua must pass over the top of the 
potential, therefore, in the limit of a very thick wall we can regard the 
bounce as a perturbation of the solution that sits on the top of the 
potential ($\phi = \phi_{\rm top}$) and linearize the equations of motion 
around this exact dS solution. To leading order, we have the equation 
for the scalar field $\phi=\phi_{\rm top}+\delta\phi$: 
\be
\Box \delta \phi \approx V'' (\phi_{\rm top}) \delta \phi,
\ee
with the geometry corrected at order ${\cal O} (\delta \phi^2)$. They analysed 
this scalar equation in the dS background, discovering that it yielded an 
eigenvalue equation for $\beta = 3V'' (\phi_{\rm top})/(8 \pi G V_{\rm top})$, 
hence a lower bound on $\beta$ for the oscillating bounce solutions, 
where $V_{\rm top} \equiv V (\phi_{\rm top})$. This eigenvalue equation
represents solutions for $\delta \phi$ that remain fully within the perturbative
r\'egime, however, does not mean there are no solutions for other values of
$\beta$, only that these will enter the r\'egime in which the nonlinear
corrections to $V''(\phi)$ become important, as described below.

We will now present an analogous calculation for the seeded decay,
however note that a BH in dS space, described by
the SdS solution, is written in `static' coordinates, 
i.e.\ there is no time-dependence in the metric. The analysis of
\cite{Hackworth:2004xb} however (and indeed the original CDL 
bounce) is performed in global -- time dependent -- dS 
coordinates. We must therefore first demonstrate that the 
unstable dS solution presents the \textit{same} eigenvalue
restrictions on $\beta$, although the eigenfunction solutions
for $\delta \phi$ may be different in this patch, due to the different
analytic continuation. We should emphasise that this is a highly nontrivial 
requirement, as the analytic continuation for the $O(4)$ instanton in the 
global patch is different to the analytic continuation for the static patch.

Lorentzian de Sitter spacetime can be represented as a hyperboloid embedding
within a 5D Minkowski spacetime.
For future discussion, we note the transformation between the 
static and global dS coordinates, denoted by $(t,r,\theta, \varphi)$ and $(T,y,\theta, \varphi)$, respectively, via the hyperboloid embedding:
\be
\beal
\sin y \sinh (HT) = &X^0 = \sqrt{1-H^2 r^2} \sinh (Ht), \\
\cos y = &X^1 = \sqrt{1-H^2 r^2} \cosh(Ht), \\
\cosh (HT) \sin y \cos \theta = &X^2 = r \cos \theta,\\
\cosh (HT) \sin y \sin \theta \sin \varphi = &X^3 = r \sin \theta \sin \varphi,\\
\cosh (HT) \sin y \sin \theta \cos \varphi = &X^4 = r \sin \theta \cos \varphi,
\eeal
\label{dslorentz}
\ee
Note that for the static coordinates, $X^1\geq0$, and $X^0\leq X^1$.
The static patch is therefore seen to be a strip of the hyperboloid,
covering less than half of the global spacetime. The analytic
continuation of the global patch as written here (which is for the 
convenience of the CDL instanton), $T \to iT_E$, while sending
$X^0\to X_E^0$, is a rotation in the $X^{0,2-4}$ coordinates,
whereas the analytic continuation of the static patch, $t\to i\tau$, 
corresponds to a continuation in the $\{X^0,X^1\}$ plane.

\subsection{Thermalons in de Sitter space}

The Wick rotated static Euclidean dS solution, $t\to i\tau$ in \eqref{dslorentz},
corresponding to $\phi=\phi_{\rm top}$ is
\be
ds^2 = (1-H^2r^2) d\tau^2 + \frac{dr^2}{(1-H^2r^2)} 
+ r^2 \left (d\theta^2 + \sin^2{\theta} d\varphi^2\right ),
\label{staticds}
\ee
where
\be
H^2 = \frac{8\pi G V_{\rm top}}{3}\,.
\ee
We now analyse the leading order deviation from this background
solution by taking a small perturbation for the scalar field, $\phi = \phi_{\rm top} 
+ \delta \phi$, and expanding the potential around its maximum:
\be
V(\phi) \approx V_{\rm top} - \frac{\beta H^2}{2} \delta \phi^2 \,.
\ee
The perturbation equation for $\delta\phi$ (assuming no angular dependence)
is then
\be
(1-H^2 r^2) \delta \phi'' + 2(1-2H^2 r^2) \frac{\delta\phi'}{r} 
+ \frac{\delta \ddot{\phi}}{(1-H^2 r^2)}+ \beta H^2\delta \phi=0,
\label{deltaphieq}
\ee
where a dot denotes the derivative with respect to the Euclidean time $\tau$.
As a first step (and mirroring the analysis in \cite{Hackworth:2004xb})
look for a static solution writing $x=Hr$\footnote{Note, $x\in[0,1]$, as opposed
to the usual range $[-1,1]$.} and $\delta \phi = \hat\phi(x)/x$.
The equation for $\hat\phi$ then has the very familiar form of a
Legendre equation
\be
(1-x^2) \hat\phi'' - 2x \hat\phi' + (2+ \beta) \hat\phi = 0,
\label{Legendre_eq1}
\ee
with solution $\hat{\phi} = P_\nu(x)$, where 
\be
\beta = (\nu-1)(\nu+2).
\ee
To guarantee the regularity of solutions of (\ref{Legendre_eq1}), we impose 
the boundary conditions: $\delta\phi' = 0$ at $r=0$, and 
$2H\delta\phi' = \beta H^2 \delta\phi$ at $r = H^{-1}$,
or
\be \displaystyle
\lim_{x \to 0} \left( \frac{\hat\phi(x)}{x} \right)' = 0, 
\qquad \hat{\phi}'(1) = \frac{2+\beta}{2} \hat{\phi}(1).
\label{boundary_first}
\ee
The second boundary condition is automatically satisfied when $\nu$ is an 
integer, since $P_\nu'(1)=\frac12\nu(\nu+1)P_\nu(1)$. However, the first 
boundary condition requires that $\nu$ be an odd integer. Note that for $\nu=1$,
$P_1(x) = x$, which simply corresponds to a constant $\delta\phi$
and $\beta=0$. This is just the statement that if the potential is flat, there
is a zero mode, so we will ignore this case. Therefore, the first non-trivial 
bounce solution with the lowest eigenvalue is obtained for $\nu = 3 \ (\beta = 10)$.

Now let us consider what happens when $\nu$ is not an integer. 
The solution for $\delta\phi$ which satisfies the boundary conditions 
at $x=0$ is the hypergeometric function
\be
\delta\phi= \delta \phi_S \equiv {}_2F_1(\frac12-\frac12\nu,1+\frac12\nu,\frac32,x^2).
\label{ds_solution}
\ee
This solution diverges at the horizon as $x\to 1$, but in the full non-linear 
system this simply implies that the solution leaves the small region around 
the top of the potential where the linearisation is valid. In the the language of 
Ref.\ \cite{Balek:2004sd}, the solution {\it undershoots} when $3<\nu<4$, 
in other words the solution has a single turning point in the region $x<1$ 
when $3<\nu<4$ (see FIG.\ 2 in \cite{Balek:2004sd}). In Ref.\ \cite{Balek:2004sd}, 
they argued that an undershoot implies that there exists a solution to the 
full non-linear equations. Therefore, there is at least one CDL instanton when
\be
\left| \frac{d^2V}{d\phi^2}\right|_{\phi = \phi_{\rm top}}
> 10 \left( \frac{8 \pi V_{\rm top}}{3 M_{\rm Pl}^2} \right).
\ee
For larger values of $\nu$, the solution oscillates in a similar way as the 
integer $\nu$ solutions
before leaving the region where the linearised solution is valid.

Now let us compare the static patch static solution to the analysis 
of \cite{Hackworth:2004xb} in global coordinates, where they obtain
the solution
\be
\delta \phi = \delta \phi_G \equiv C_N^{3/2}(\cos y),
\ee
where $y$ is the ``radial'' coordinate of the global dS four-sphere scaled
by the dS curvature scale $H$, and is actually more correctly interpreted
as an angular coordinate (see \eqref{dslorentz}). $C_N$ is a Gegenbauer 
polynomial, with $\beta = N(N+3)$ for integer $N$.
Superficially this looks as if it gives the same results as ours, but in 
\cite{Hackworth:2004xb} the integer $N$ corresponds to the number of
oscillations of the solution around the top of the potential, so even $N$
corresponds to an oscillation from true (or false) vacuum back to true (false)
vacuum, and odd $N$ gives the phase transition of false to true (or v.v.).
Comparing the eigenvalues $N = \nu-1$, hence {\it odd} $\nu$ corresponds
to {\it even} $N$ -- i.e.\ no phase transition! It would seem therefore on a 
cursory inspection that the static patch is problematic, and not giving equivalent 
results; indeed, taking the lowest allowed $\nu-$value $\nu=3$ would indicate 
a lower bound on $\beta$ of $\beta \gtrsim 10$, or 
$|V'' (\phi_{\rm top})| \gtrsim 10 H^2$ as opposed to $4H^2$. 

The resolution of this problem is readily 
found by returning to the original work in the thin wall limit \cite{Gregory:2013hja}, 
where it was noted that the CDL instanton in the static patch is time dependent.
Reinstating the time dependence in \eqref{deltaphieq}, 
we note that the periodicity of $\tau$, $\tau \sim \tau + 2\pi/H$, implies that 
$\delta\phi$ must take the form $\delta \phi = e^{i\mu H\tau} \hat\phi(x)/x$, 
where $\mu$ is an integer. Substituting this into (\ref{deltaphieq}), one 
obtains the associated Legendre equation:
\be
(1-x^2) \hat\phi'' - 2x \hat\phi' + \left (2+ \beta 
- \frac{\mu^2}{1-x^2}\right) \hat\phi = 0
\ee
with the same eigenvalues $\beta = (\nu-1)(\nu+2)$, and solution
\be
\delta \phi = P_\nu^\mu(x) e^{i\mu H\tau}.
\ee
The presence of the time dependence modifies the boundary condition 
for $\hat\phi$, that now has to vanish at $x=1$ as well as at $x=0$; this
is a crucially different boundary condition from (\ref{boundary_first}). 
The former requirement is satisfied for nonzero $\mu$, and the latter for 
$\mu+\nu$ an odd integer. In particular, we may now include {\it even} 
values of $\nu$, the lowest of which yields the required bound 
$\beta \gtrsim 4$ or $|V'' (\phi_{\rm top})| \gtrsim 4H^2$, that is now fully
consistent with \cite{Hackworth:2004xb}.

It is interesting to compare these static patch Euclidean solutions to the 
global patch Euclidean solutions in \cite{Hackworth:2004xb}. As we noted,
in the Lorentzian dS space, the static patch covers only the inside of the 
cosmological horizon, and less than half of the dS hyperboloid \eqref{dslorentz}.
Further, the analytic continuation of the global patch is a rotation affecting
the embedded time and $2,3,4$-coordinates, whereas the static patch 
continuation is a combination of the embedding time and the remaining spatial 
coordinate. Performing this transformation, $T \to iT_E$, $X^0\to X^0_E$, 
and $t\to i\tau$, on each side of \eqref{dslorentz},
\be
\beal
\sin y \sin (HT_E) = &X^0_E = \sqrt{1-H^2 r^2} \sin (H\tau), \\
\cos y = &X^1 = \sqrt{1-H^2 r^2} \cos(H\tau), \\
\cos (HT_E) \sin y \cos \theta = &X^2 = r \cos \theta,\\
\cos (HT_E) \sin y \sin \theta \sin \phi = &X^3 = r \sin \theta \sin \varphi,\\
\cos (HT_E) \sin y \sin \theta \cos \phi = &X^4 = r \sin \theta \cos \varphi,
\eeal
\label{dseuclid}
\ee
we now see that something interesting has happened.
In the static patch, we typically choose the periodicity of
Euclidean time to render the space regular, i.e.\ $\tau \sim \tau + 2\pi/H$
as used above. But now this recovers the full range of the
Euclidean $X^1$ coordinate, $X^1 \in [-1,1]$, whereas in the 
Lorentzian section, $X^1\geq0$. This recovery means that 
in fact both the `global' and the `static' Euclidean continuation of dS
cover the full space, and have the same, $S^4$, topology!
This is a somewhat surprising result since the Lorentzian coordinate patches
are very different, as is the analytic continuation. Using \eqref{dseuclid}, 
the coordinate transformation from the static patch to global one is obtained as 
\be
\sin (H T_E) = \frac{\sqrt{1-H^2 r^2} \sin H\tau}
{\sqrt{\cos^2 H\tau + H^2 r^2 \sin^2 H\tau}}\;\;\;,\quad
\cos y = \sqrt{1-H^2r^2} \cos H\tau.
\label{trans_G_S}
\ee

\begin{figure}[t]
\centering
\includegraphics[width=0.3\linewidth]{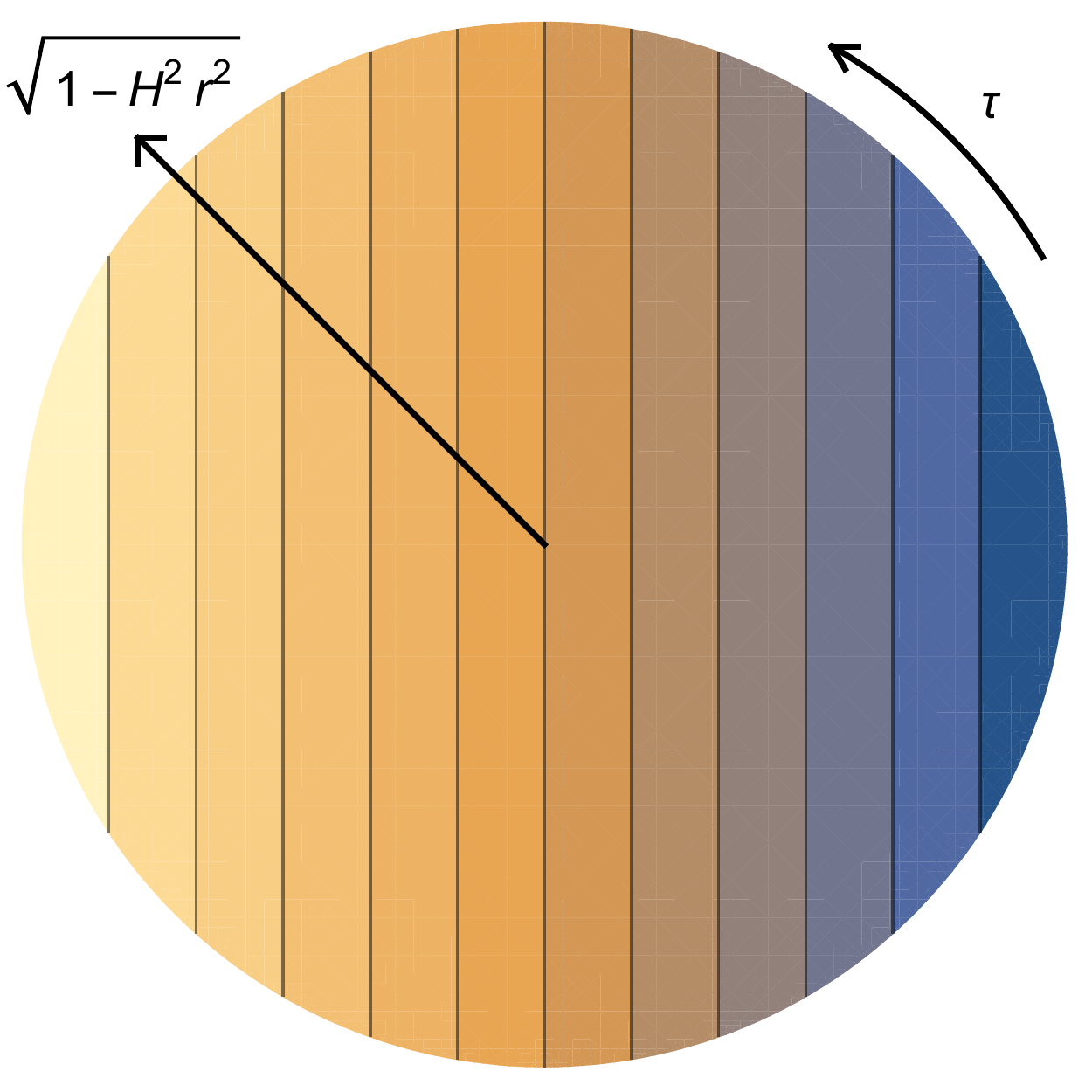}\hskip 2cm
\includegraphics[width=0.5\linewidth]{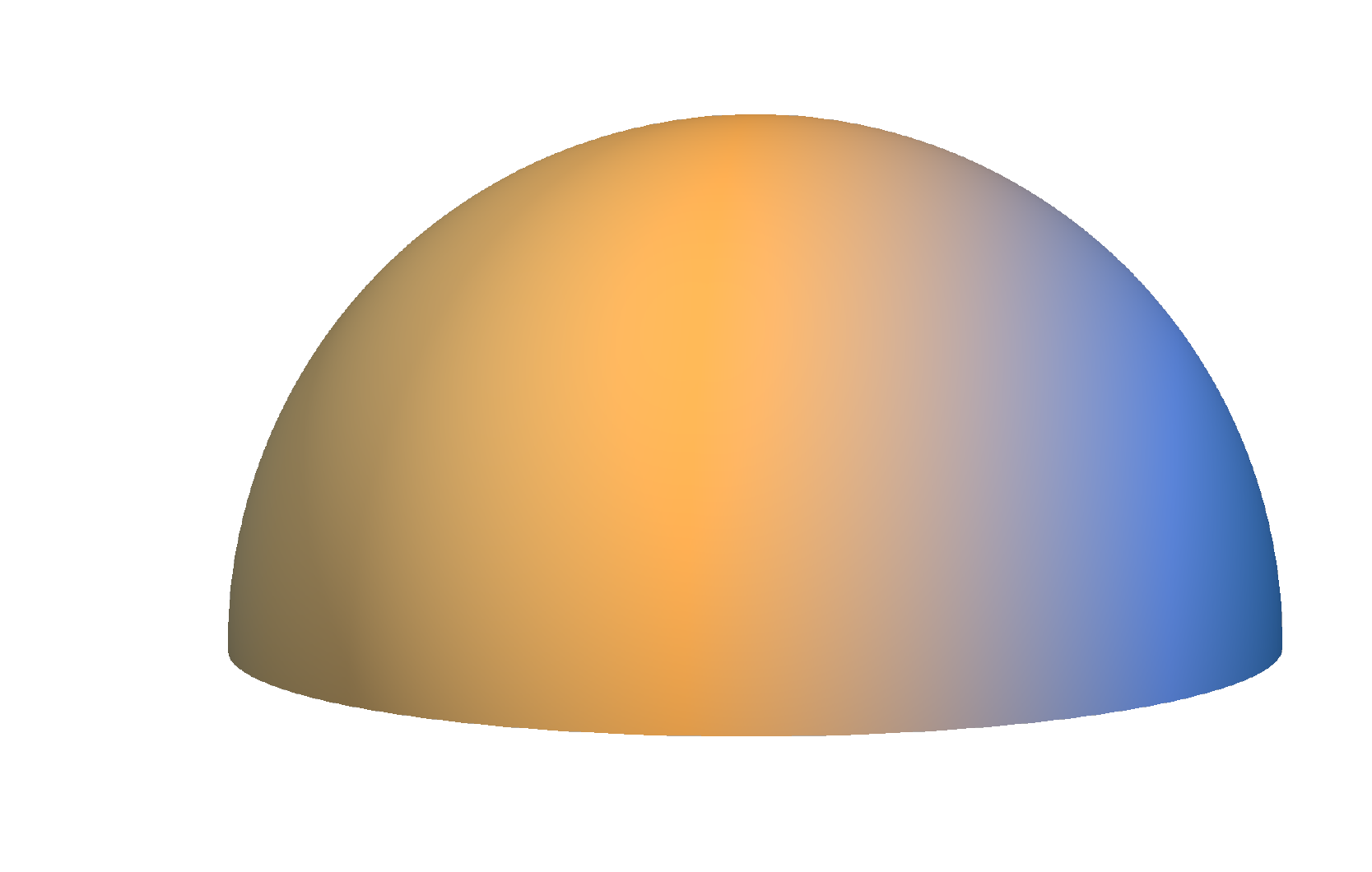}
\caption{The lowest lying eigenfunction ($\beta = 4$), shown both as a projection
onto the plane, and also with the hemispherical geometry of the $\tau - r$ 
$(\tilde{\tau}-x)$ section
that makes a stronger link to the global analytic continuation (although here the
radius of the suppressed $S^2$ is actually zero on the equator!)}
\label{fig:lowest}
\end{figure}

\begin{figure}[h]
\centering
\includegraphics[width=0.3\linewidth]{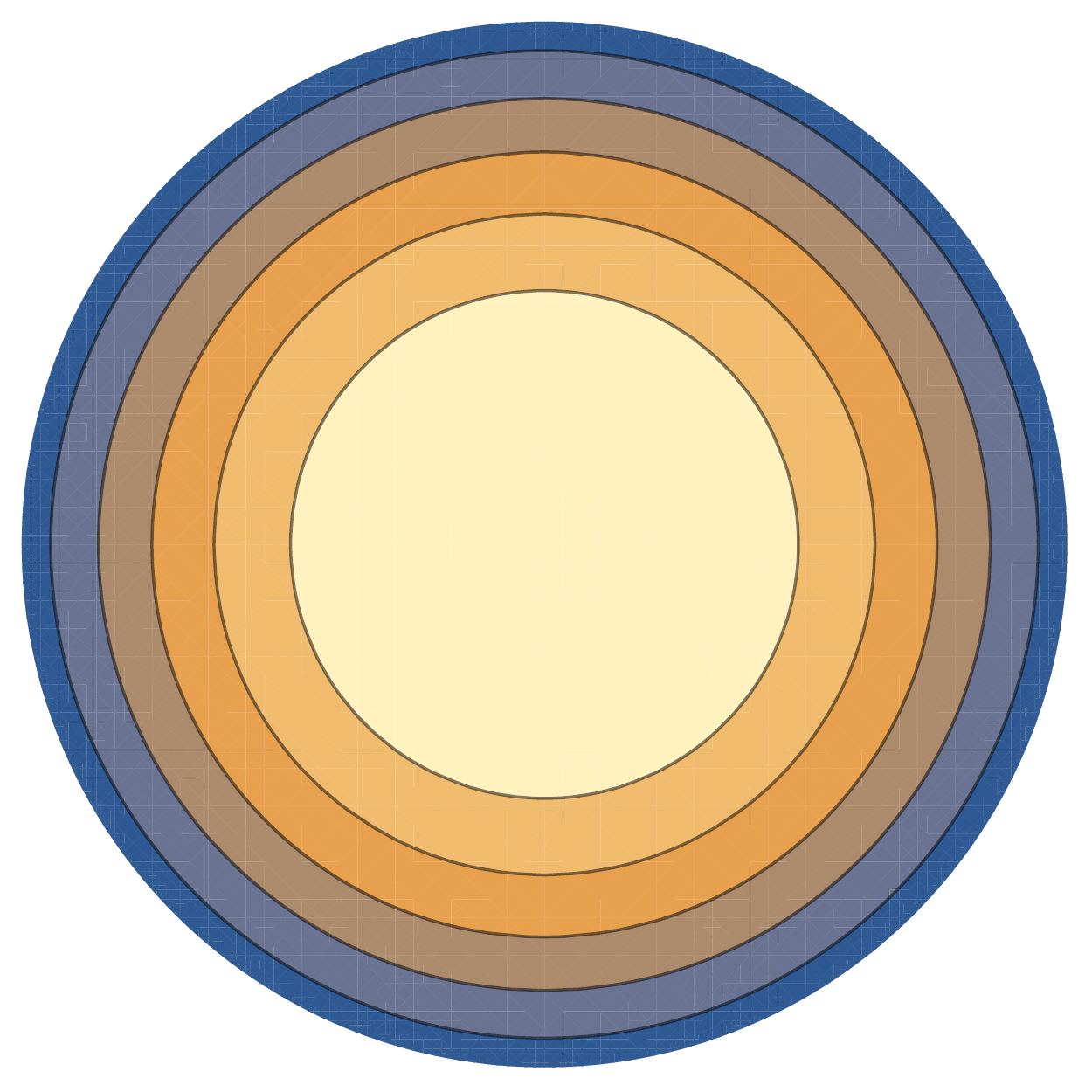}
\includegraphics[width=0.3\linewidth]{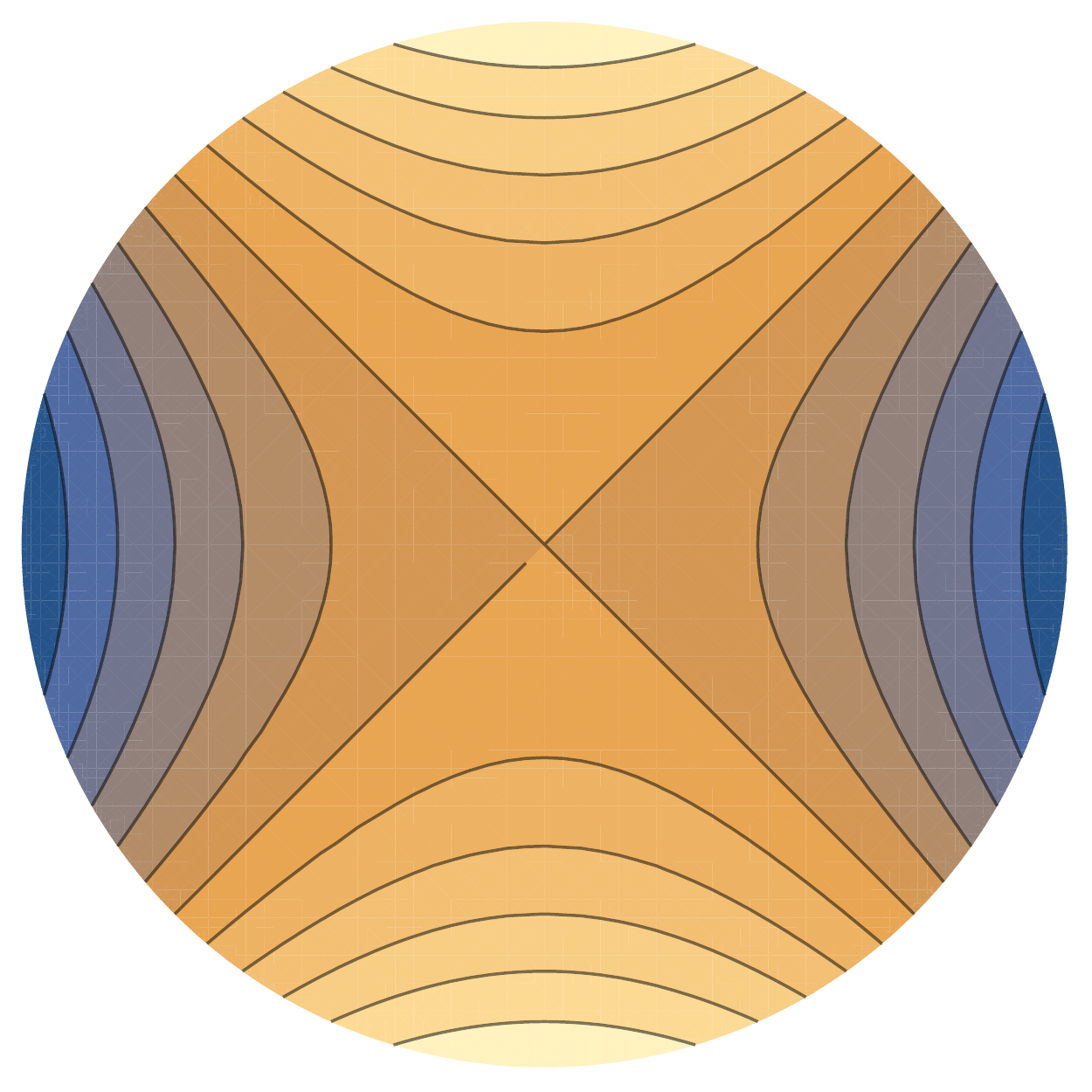}
\includegraphics[width=0.3\linewidth]{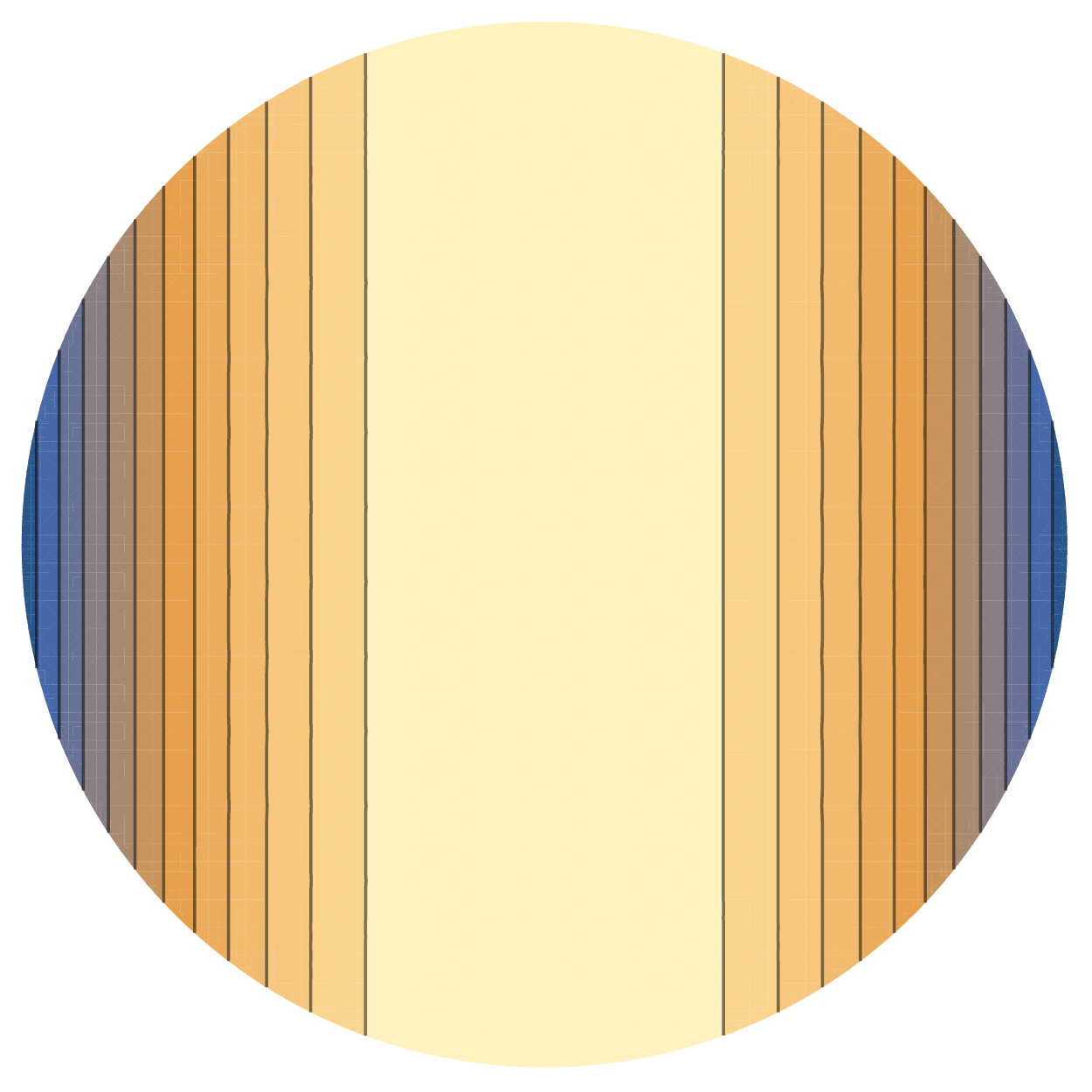}
\caption{The first harmonic eigenfunction ($\beta = 10$). On the left and in 
the middle are the two eigenfunctions, $P_3^0(x)/x$ and 
$P_3^2(x)$cos$(2\tilde{\tau})$. On the right is the combination of the two 
eigenfunctions that represents the transformed Gegenbauer solution 
from \cite{Hackworth:2004xb}.}
\label{fig:next}
\end{figure}

Now that we see that the two different coordinate descriptions are just 
alternative ways of representing the same manifold, we expect that the 
scalar solutions presented in \cite{Hackworth:2004xb} in terms of 
Gegenbauer polynomials will be expressible in terms of the Legendre
functions we have obtained in the static patch by the completeness
properties of spherical harmonics on the sphere.
Explicitly, the eigenfunctions of \cite{Hackworth:2004xb} are
$\delta\phi = \delta \phi_G \equiv C_N^{3/2} (\cos y)$, (where the $C_N$ 
are the Gegenbauer polynomials), and, writing $x=Hr$ and $\tilde{\tau}=H\tau$, 
our static patch eigenfunctions are
\be
\delta \phi = \delta \phi_S \equiv
\sum_{l = \left [ \frac{\nu}{2}\right]} ^{\nu-1} 
A_{\nu}^l \frac{P_{\nu}^{2 l -\nu+1}(x) }{x} \cos\left((2 l -\nu+1) \tilde{\tau} \right)
\ee
where the square brackets in the lower limit stand for {\sl integer part},
and $A_{\nu}^l$ is an constant. To give a few examples, let us compare the 
lowest two eigenfunctions:
\be
\beal
\nu=2 \; ; 
\beta =4 \qquad
&\begin{cases}
\delta \phi_G = C_1^{3/2} (\cos y) =  3 \cos y & \\
\delta \phi_S = A_2^1 \cos \tilde{\tau} P_2^1 (x) /x
= -3 A_2^1 \sqrt{1-x^2} \cos \tilde{\tau}, &
\end{cases}
\label{beta_4_del_phi}
\eeal
\ee
\be
\beal
\nu = 3\; ;
\beta =10 \qquad
&\begin{cases}
\delta \phi_G = C_2^{3/2} (\cos y) = 3(5 \cos^2y-1)/2 &  \\
\delta \phi_S = A_3^1 P_3^0(x)/x + A_3^2 \cos(2\tilde{\tau}) P_3^2 (x) /x  \\
~~~~~ = - A_3^1 (3 - 5 x^2)/2 + 15 A_3^2 (1-x^2) \cos(2\tilde{\tau}). &
\end{cases}
\label{beta_10_del_phi}
\eeal
\ee
Using (\ref{trans_G_S}), we see immediately that $\delta \phi_G = \delta \phi_S$ 
for $A^1_2=-1$, in (\ref{beta_4_del_phi}), corresponding to the solution
\be
\delta \phi \propto \sqrt{1-x^2} e^{i\tilde{\tau}},
\ee
depicted in FIG.\ \ref{fig:lowest}. Setting $A_3^1 = -6 A_3^2=-3/2$ in \eqref{beta_10_del_phi}, one obtains $\delta \phi_S = 3 (5 \cos^2 y -1)/2$ for 
$\beta = 10$, and both $\delta \phi_S$ and $\delta \phi_G$ give an oscillating 
bounce with false$\to$true$\to$false (or v.v.) as is shown in FIG.\ \ref{fig:next}. 
Thus, the analysis in the static and global patches are explicitly consistent,
at least at the level of the eigenvalue analysis of the perturbation of the scalar
field. It would be interesting to consider what  might happen beyond the linear level,
as the thin wall CDL instanton has a different periodicity of Euclidean time, and a 
conical deficit at the cosmological horizon.

Finally, we briefly mention the physical interpretations for $\delta \phi_S$ 
and $\delta \phi_G$. According to the arguments in 
\cite{Gomberoff:2005je,Brown:2007sd,Masoumi:2012yy,Ai:2018rnh}, a static solution in the 
static patch may be regarded as a thermal transition (thermalon), although the 
solution in the global patch that is no longer ``static'' has the interpretation of
a quantum tunneling event. Based on this discussion, we would argue that 
the static patch $\delta \phi_S$ may describe a thermal transition inside 
the cosmological horizon. (Extension to SdS space is provided in 
Sec.\ \ref{sec_SdS_numerical}.) In addition, a thermal system is an ensemble state of energy and a thermalon would be suppressed by the Boltzmann factor $e^{-\Delta F/T}$, where $\Delta F$ is the change of free energy and $T$ is a temperature. In our situation, the internal energy vanishes due to the Hamiltonian constraint and $\Delta F = - T \Delta {\cal S}$, which leads to the entropic Boltzmann factor (\ref{boltzmann}). This picture is based on an excitation transition and a metastability of initial state, that is guaranteed by the existence of one and only one negative mode, may be not necessary. Indeed, it was shown that a bounce solution with $N$ oscillations has $N$ negative modes in dS background \cite{Lavrelashvili:2006cv,Battarra:2012vu}. We do not have any conclusive argument regarding the role of the negative modes in thermal transition and it is an interesting open question.

\subsection{Thermalons near the Nariai limit}

Having verified that the static patch analysis in Euclidean dS, which 
may be viewed as  the $M\to0$ limit of SdS, gives equivalent results
to the global $O(4)$ analysis, we now present another limit in which we
can analytically explore bounce solutions: the Nariai limit \cite{Nariai:1,Nariai:2}.
The SdS spacetime has two horizons: the BH and the cosmological event 
horizons, whose radii are denoted by $r_h$ and $r_c$ respectively, 
that translate into two bolts at each end of the Euclidean section. 
As the mass of the BH is increased, these two horizons move towards each 
other, until at $G MH = 1/\sqrt{27}$, the horizons merge at
$r_h=r_c=1/\sqrt{3}H$. This is the Nariai limit. 
Near this limit, the $\tau-r$ metric functions can be approximated by 
quadratic functions, allowing once again a simple perturbation equation for 
the scalar field with analytic solutions for some eigenvalues. This time 
however, we do not expect to have time dependent solutions for our scalar
field, as the preferred instantons for large BHs are static (at least in the
thin wall case) \cite{Gregory:2013hja}.

First, we write down the metric near the Nariai limit:
\begin{equation}
ds^2 = 3 H^2 (r_\epsilon^2 - \bar{r}^2) d\tau^2 
+ \frac{d\bar{r}^2}{3 H^2 (r_\epsilon^2 - \bar{r}^2)} + R^2 d\Omega^2,
\end{equation}
where $r_\epsilon \equiv (r_c - r_h)/2$, $R \equiv (\sqrt{3} H)^{-1}$, 
and $\bar{r} \equiv r- R$. Changing the scale of coordinates as before,
$x = \bar{r} /\epsilon$ and $\tilde{\tau} = H\tau$, we finally obtain
\begin{equation}
ds^2 = 3 r_\epsilon^2 (1-x^2) d\tilde{\tau}^2 
+ \frac{dx^2}{3 H^2 (1-x^2)} + R^2 d\Omega^2,
\end{equation}
for which the equation of motion of a massless scalar field is
\begin{equation}
(1-x^2) \delta \phi'' -2 x \delta \phi' + \frac\beta3\delta \phi \approx 0.
\end{equation}
Thus, once again, the perturbation equation for the scalar reduces to 
a Legendre equation, although this time without any transformation on 
$\delta\phi$. Note also that now the equation is valid in 
the range $x\in[-1,1]$, as the range of the $\bar{r}$ coordinate is 
$[-\epsilon,\epsilon]$. The solution is
\begin{equation}
\delta\phi = P_\nu(x),
\label{nariai_solution}
\end{equation}
where $\beta = 3\nu(\nu+1)$, and the boundary condition at each horizon is simply that 
\be
\delta \phi' (\pm 1) = \pm \frac\beta6 \delta\phi (\pm1) 
= \pm \frac{\nu(\nu+1)}{2} \delta\phi (\pm1),
\ee
which are satisfied for all $\nu$. The lowest eigenfunction with $\nu=1$,
corresponds to a solution interpolating between $T\to F$ from the black
hole to the cosmological horizon. This solution has $\beta=6$, and we label
it as a fundamental mode, or $k=1$. 

Let us now consider what happens to this fundamental mode as we move
away from the Nariai limit. Naturally, the analytic approximation above becomes
less valid, as the cubic nature of the Newtonian potential comes into play.
Qualitatively however, we expect that there will still exist a fundamental solution
interpolating from `$T$' at $r_h$ to `$F$' near $r_c$. As we gradually switch off
the mass, what remains will be a solution interpolating from `$T$' at $r=0$ 
to `$F$' at $r_c=1/H$; this however is none other than the solution 
$P_3^0(x)/x$ illustrated on the left in FIG.\ \ref{fig:next}. In other words,
our fundamental mode has continued to the first harmonic in the static patch,
moreover, the value of $\beta$ has increased from $\beta_N = 6$ to 
$\beta_{dS} = 10$. We will therefore be looking for confirmation of this
drift of $\beta$ with $M$ in the full numerical calculation.

\section{Oscillating bounce in general SdS space}
\label{sec_SdS_numerical}

In this section, we will numerically investigate oscillating bounce solutions 
in the general SdS background. We solve the full Einstein-scalar equations,
thus taking account of the gravitational backreaction of the bounce on the 
background spacetime. We will show that the mass of a seed BH, $M_+$, 
and the curvature of the potential at the top of the barrier, $\beta$, determine 
the maximum number of possible oscillations of the bounce. An oscillating 
bounce with a large number of oscillations has a super-thick bubble wall, 
which can be regarded as an up tunneling, such as the HM transition, around 
a BH. An interpretation of the oscillating bounce around a BH will be discussed 
in the next section in more detail.

\subsection{Methodology}

\begin{figure}[h]
\centering
\includegraphics[width=0.7\textwidth]{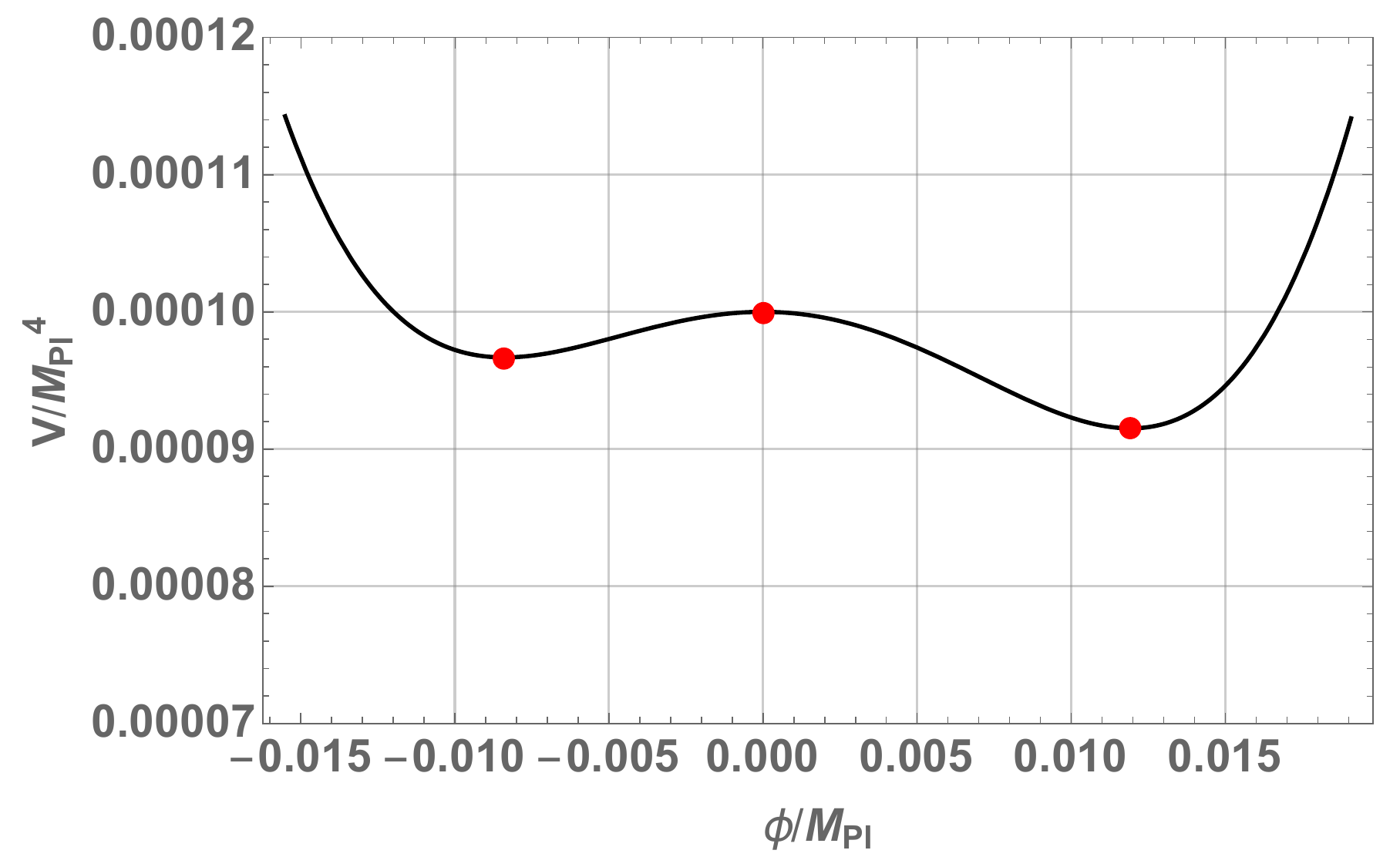}
\caption{Plot of the effective potential $V(\phi)$ with $V_{\rm top} 
= 10^{-4} M_{\text{Pl}}^4$, $\beta = 250$, $g= 1/\sqrt{8}$, and 
$v= 0.01 M_{\text{Pl}}$. Red points show the false vacuum, true vacuum, 
and the top of barrier.
}
\label{potential}
\end{figure}

For simplicity, we use the following effective potential throughout the manuscript
\begin{equation}
V (\phi) = \beta H^2 \left( - \frac{1}{2} \phi^2 - \frac{g}{3 v} \phi^3 
+ \frac{1}{4 v^2} \phi^4 \right) + V_{\rm top},
\end{equation}
where $H^2 = (8 \pi G/3) V_{\rm top}$ and $\beta = |V'' (\phi_{\rm top})|/H^2$. 
The potential is constructed so that it's local maximum, $V_{\rm top}$ is at
$\phi=0$ and the true and false vacuum states are at $\phi = 
\phi_T \equiv \frac{v}{2} (g + \sqrt{4 + g^2})$ and $\phi = 
\phi_F \equiv - \frac{v}{2} (\sqrt{4 + g^2} - g)$, respectively. 
We take $V_{\rm top} = 10^{-4} M_{\text{Pl}}^4$, $\beta \leq 250$,
$g= 1/\sqrt{8}$, and $v= 0.01 M_{\text{Pl}}$ throughout the manuscript 
(FIG.\ \ref{potential}).
We consider a phase transition from a uniform configuration of the 
scalar field $\phi = \phi_F$ with a seed BH of mass $M_+$ to a spherical 
and inhomogeneous configuration. In the background of a BH, $O(4)$ 
symmetry is broken to $O(3)\times U(1)$, and in the following we 
investigate oscillating bounces preserving this symmetry.

To construct a static bounce around a BH, we assume a spherically 
symmetric geometry whose generic (Euclidean) metric has the form of
\begin{equation}
ds^2 = f(r) e^{2 \delta(r)} d\tau^2 + \frac{dr^2}{f(r)} + r^2 (d\theta^2 
+ \sin^2{\theta} d\varphi^2),
\end{equation}
where the metric function $f(r)$ is defined as
\begin{equation}
f= 1- \frac{2G \mu (r)}{r}.
\end{equation}
We have to numerically solve the Einstein and scalar equations simultaneously 
in order to determine the functions $\mu(r)$, $\delta(r)$ and $\phi(r)$:
\bea
&& f \phi'' +f' \phi' + \frac{2}{r} f \phi' + \delta' f \phi' - V_{\phi} = 0,
\label{FEQ1} \\
&& \mu' = 4 \pi r^2 \left( \frac{1}{2} f \phi' {}^2 + V \right),
\label{FEQ2} \\
&& \delta' = 4 \pi G r \phi' {}^2,
\label{FEQ3}
\eea
where a prime denotes the derivative with respective to $r$. 
Note that we can eliminate $\delta'$ from \eqref{FEQ1} using
\eqref{FEQ3}, meaning that we solve first for $\phi$ and $\mu$, then
recover $\delta$ from \eqref{FEQ3}. In the following calculation we use 
the tortoise coordinate $r^{\ast}$ instead of $r$ since the behavior of 
the scalar field near a horizon is clearer in this coordinate:
\begin{equation}
dr^{\ast} \equiv \frac{dr}{f(r)}.
\end{equation}
Using the tortoise coordinate, \eqref{FEQ1} and \eqref{FEQ2} reduce to
\begin{align}
&\frac{d^2 \phi}{dr^{\ast} {}^2} + f (r) \frac{2}{r} \frac{d \phi}{dr^{\ast}} 
+ 4 \pi G f^{-1} (r) r \left( \frac{d \phi}{dr^{\ast}} \right)^3 - f (r) V_{\phi} = 0,
\label{kame1}\\
&\frac{d \mu}{dr^{\ast}} = 4 \pi r^2 \left( \frac{1}{2} \left( \frac{d \phi}{dr^{\ast}} \right)^2 
+ f (r) V \right).
\label{kame2}
\end{align}
In order for the third term in \eqref{kame1} to be regular at the BH horizon 
and at the cosmological horizon ($r^{\ast} \to \pm \infty$),  we have to 
impose boundary conditions:
\begin{equation}
\frac{d \phi}{dr^{\ast}} \to 0 \ \text{for} \  r^{\ast} \to \pm \infty.
\label{condition1}
\end{equation}
We additionally impose
\begin{align}
\lim_{r^{\ast} \to - \infty} \mu (r^{\ast}) &= \mu_-,\\
\displaystyle \lim_{r^{\ast}  \to \infty} \phi (r^{\ast}) &= \phi_F,
\label{boundary_cond_phif}
\end{align}
where $\mu_-$ is a constant to be determined freely. Note that $\mu_-$ is 
not the ``mass'' of the remnant BH per se, as it includes the contribution from 
the cosmological constant to the potential $f$. The metric function in the 
vicinity of the horizon is matched with that of the SdS metric using the 
boundary condition (\ref{condition1}), which leads to
\begin{align}
&f(r_h) = 1- \frac{2GM_-}{r_h} -\frac{8 \pi G}{3} V(\phi_h) r_h^2 
= 1- \frac{2G \mu_-}{r_h} \ \text{for} \ r^{\ast} \to -\infty,\\
& \longrightarrow \mu_- = M_- + \frac{4 \pi}{3} r_h^3 V (\phi_h),
\end{align}
where $r_h$ denotes the radius of the remnant BH horizon and 
$\displaystyle \phi_h \equiv \lim_{r^{\ast} \to -\infty} \phi$. 

On the other hand, one can obtain the total `mass' function inside the 
cosmological horizon as
\begin{equation}\displaystyle
\mu_+ \equiv \lim_{r^{\ast} \to \infty} \mu (r^{\ast}).
\end{equation}
As before, this is a combination of black hole and cosmological constant 
terms:
\begin{equation}
\mu_+ = M_+ + \frac{4 \pi}{3} r_c^3 V(\phi_F).
\end{equation}
For our solutions, we require that $\mu_+$ is the same
before and after the phase transition, giving a 
one-to-one correspondence between $M_-$ and $M_+$ 
for a given number of oscillations $k$.
\begin{figure}[t]
\centering
\includegraphics[width=0.9 \textwidth]{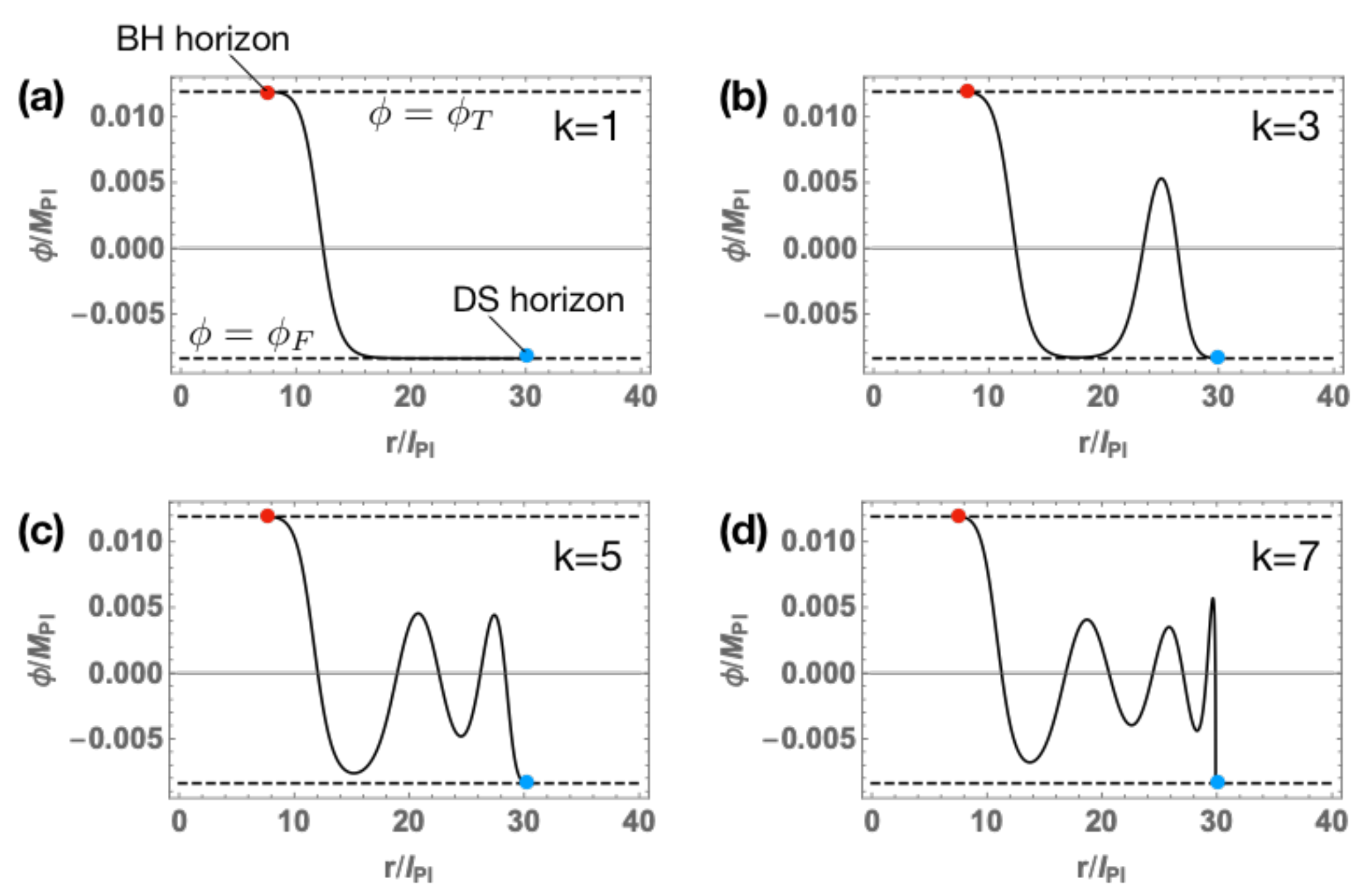}
\caption{Bounce solutions for $k=1, 3, 5,$ and $7$ with $\beta = 250$ 
and $(G^3 V_{\text{top}})^{1/2}\mu_- =0.04$. Red and Blue points denote 
the BH and cosmological horizons, respectively.
}
\label{solutions}
\end{figure}
\begin{figure}[b]
\centering
\includegraphics[width=0.8\textwidth]{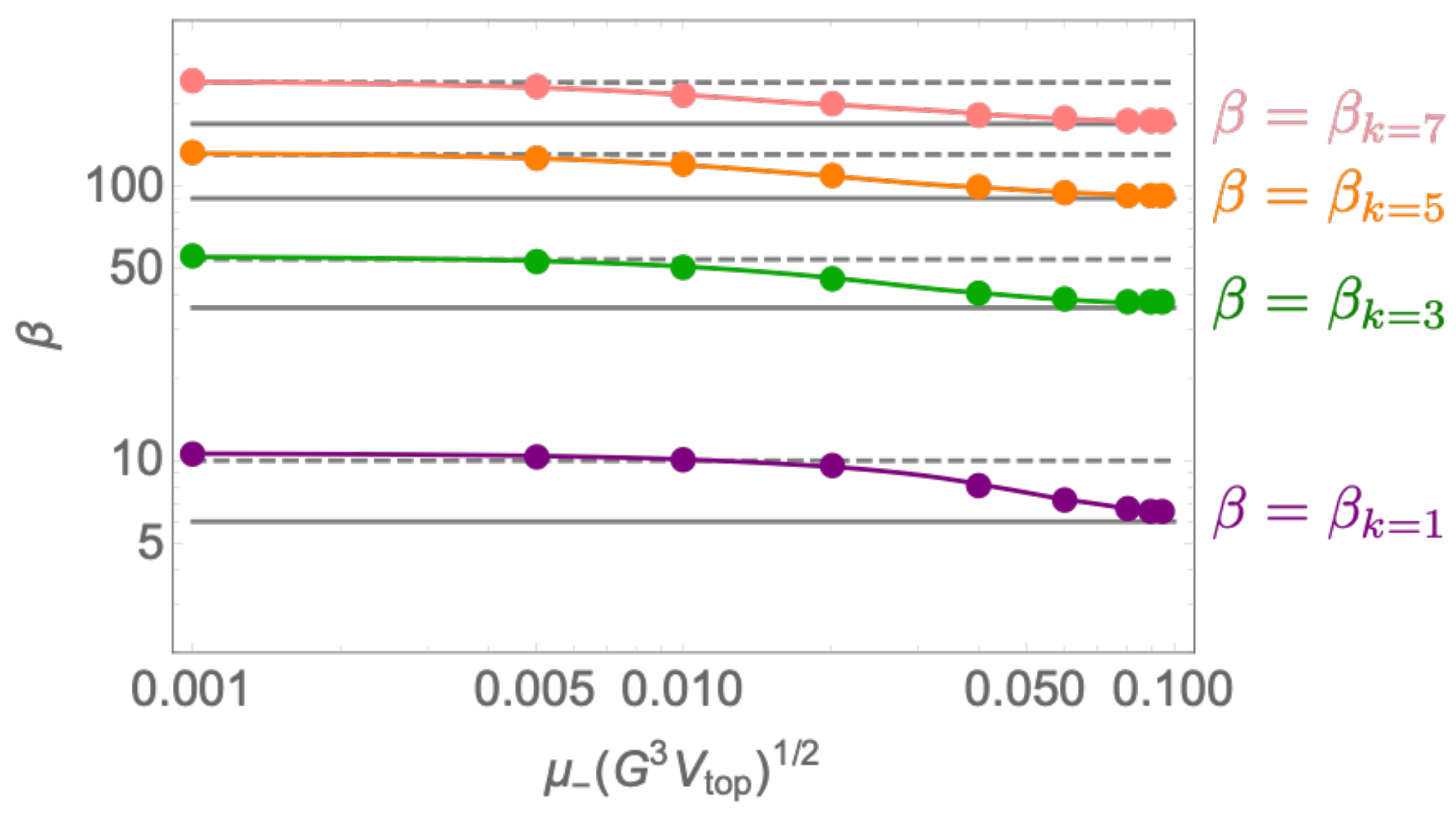}
\caption{Plot of $\beta_{k=1,3,5,7}$ above which there exist the oscillating 
bounces with $k$ oscillations. Solid and dashed gray lines show the values 
of $\beta_k$ analytically obtained in the Nariai and dS limits, respectively. 
Both are well consistent with the numerical results.
}
\label{k_beta}
\end{figure}

Using a shooting method, we numerically calculated the bounce solutions,
finding oscillating solutions in which $\phi$ crosses the top of the potential 
barrier $k$-times ($k\geq1$) (FIG.\ \ref{solutions}). Tuning the value of 
$\phi_h$, we can obtain solutions with $k=1$, $3$, $5$, and $7$ in the 
case of $\beta = |V'' (\phi_{\rm top})|/H^2 = 250$. It is interesting to consider 
the reason why $\phi$ can have such an oscillatory behavior. To construct a 
bounce solution, we have to solve the field equation (\ref{kame1}), which can
be thought of as classical motion of a particle in the inverted potential via
the usual Coleman interpretation. The field $\phi$ starts to roll down from 
$\phi \simeq \phi_T$ at $r= r_h$ aiming for $\phi= \phi_F$. If the
field does not have ``enough energy'' to reach $\phi_F$ in the intermediate 
region ($r_h < r < r_c$) it will oscillate, however, as the cosmological horizon
is approached, the potential slope becomes very shallow due to $f(r) \to 0$ 
in the vicinity of $r \simeq r_c$, thus allowing $\phi \to \phi_F$ in the vicinity 
of the cosmological horizon $r^{\ast} \to + \infty$.

\subsection{Results}

\begin{figure}[h]
\centering
\includegraphics[width=0.85 \textwidth]{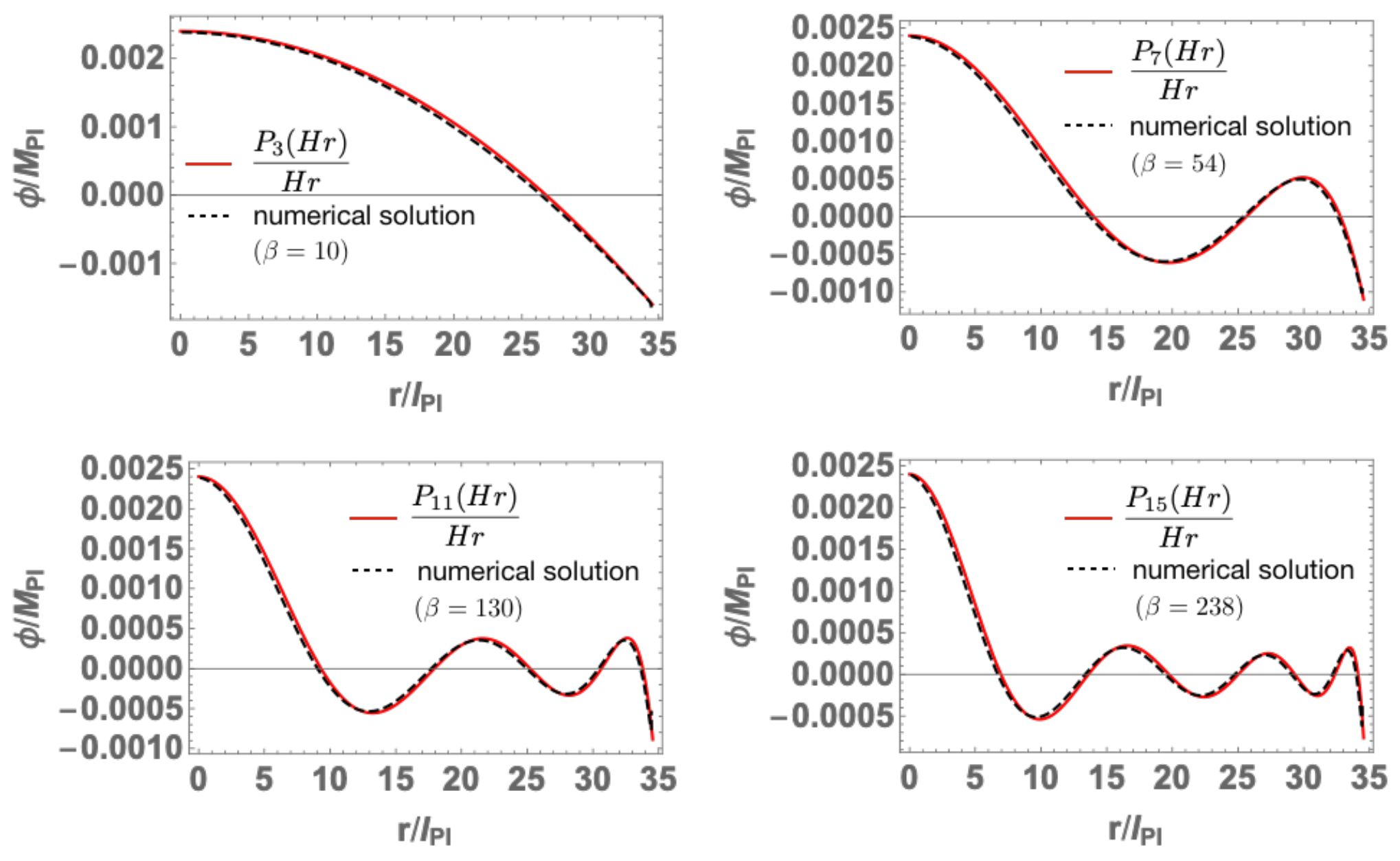}
\caption{Comparison between the numerically obtined oscillating bounce 
solutions (black dashed) and analytic solutions in the dS limit (red solid).
}
\label{comp_ds}
\includegraphics[width=0.85 \textwidth]{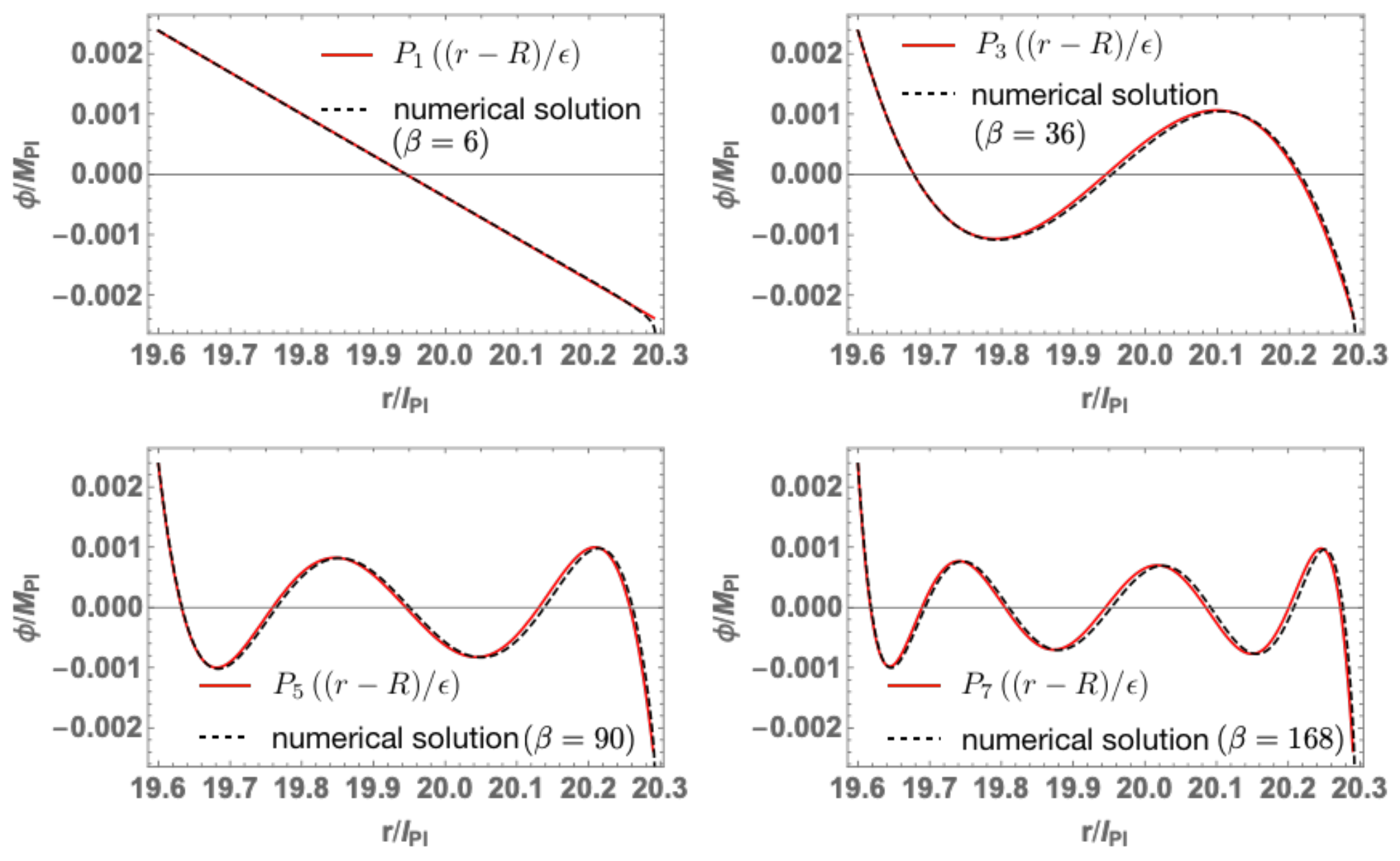}
\caption{Comparison between the numerically obtined oscillating bounce 
solutions (black dashed) and analytic solutions in the Nariai limit (red solid).
}
\label{comp_na}
\end{figure}

We show the dependence of the maximum number of $k$ on $\beta$ and 
$\mu_-$ in FIG.\ \ref{k_beta}. When the curvature at the top of the potential, 
$\beta$, is much smaller than unity, the only allowed solution is the HM 
bounce, and oscillating and monotone bounces are not allowed. The 
oscillating bounce with $k$-times crossing is allowed for $\beta \geq \beta_k$ 
as is shown in FIG.\ \ref{k_beta}. The HM bounce always allowed in principle, 
but the bounce solutions with $k= 1, 3, 5...$ are limited by $\beta \geq \beta_k$. 
The numerically obtained $\beta_k$ is pleasingly consistent with the analytically 
obtained lower bounds $\beta_k \geq 2k (2k+3)$ (gray dashed lines in FIG.\ 
\ref{k_beta}) and $\beta_k \geq 3 k (k+1)$ (gray solid lines in FIG.\ \ref{k_beta}) 
in the dS limit and Nariai limit, respectively. We also compared the analytic 
solutions obtained by the background geometries, (\ref{ds_solution}) and 
(\ref{nariai_solution}), and the numerical solutions. Then we confirmed that 
numerical solutions are in good agreement with the analytic solutions expressed 
by the Legendre equation (see FIG.\ \ref{comp_ds} and FIG.\ \ref{comp_na} 
for the dS and Nariai limit, respectively). We should note that we relaxed the 
boundary condition (\ref{boundary_cond_phif}) in the numerical results 
shown in FIG.\ \ref{comp_ds} and $\ref{comp_na}$. 

\begin{figure}[h]
\centering
\includegraphics[width=0.8 \textwidth]{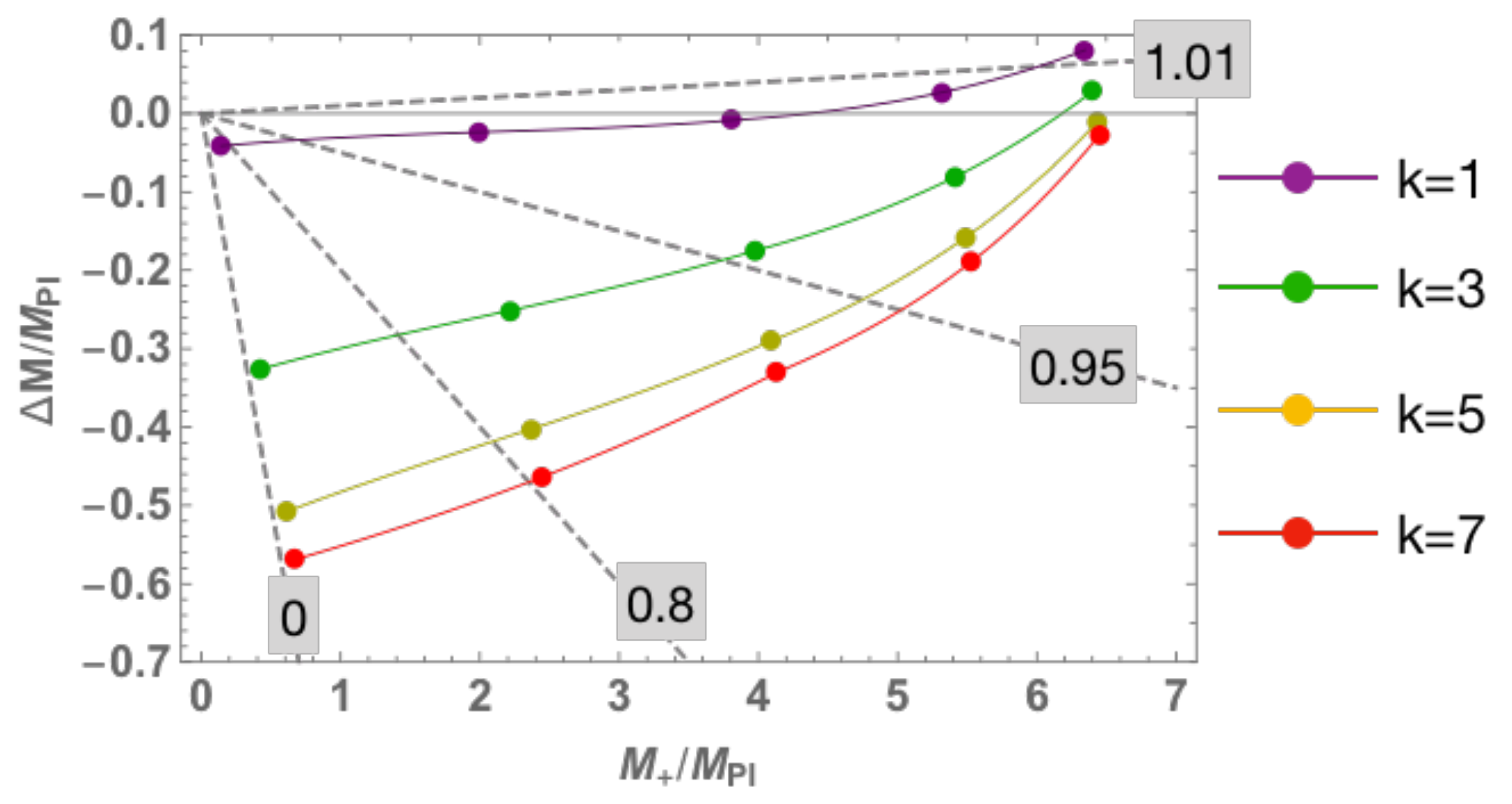}
\caption{Plot of the mass change $\Delta M = M_--M_+$ with $\beta = 250$ 
due to the phase transition. Gray dashed lines show $M_-/M_+ = 0, 0.8, 0.95,$ 
and $1.01$. 
The Nariai limit is at $M_+ \simeq 6.7 M_{\text{Pl}}$ in our setup.
}
\label{mass_change}
\end{figure}

We also investigate the change of BH mass before and after the phase 
transition for $k=1, 3, 5$, and $7$ (FIG.\ \ref{mass_change}). Since the 
oscillating bounce with a higher value of $k$ gives a thicker bubble wall, 
a larger amount of BH mass is consumed to balance the increment of the 
total vacuum energy inside the cosmological horizon. However, there is 
not enough ``room'' to oscillate and to excite the vacuum energy in the 
Nariai limit, and so the change of the BH mass is suppressed in the limit.

\section{Hawking-Moss and Oscillating bounces around a BH}
\label{sec_HM}

\begin{figure}[h]
\centering
\includegraphics[width=0.8 \textwidth]{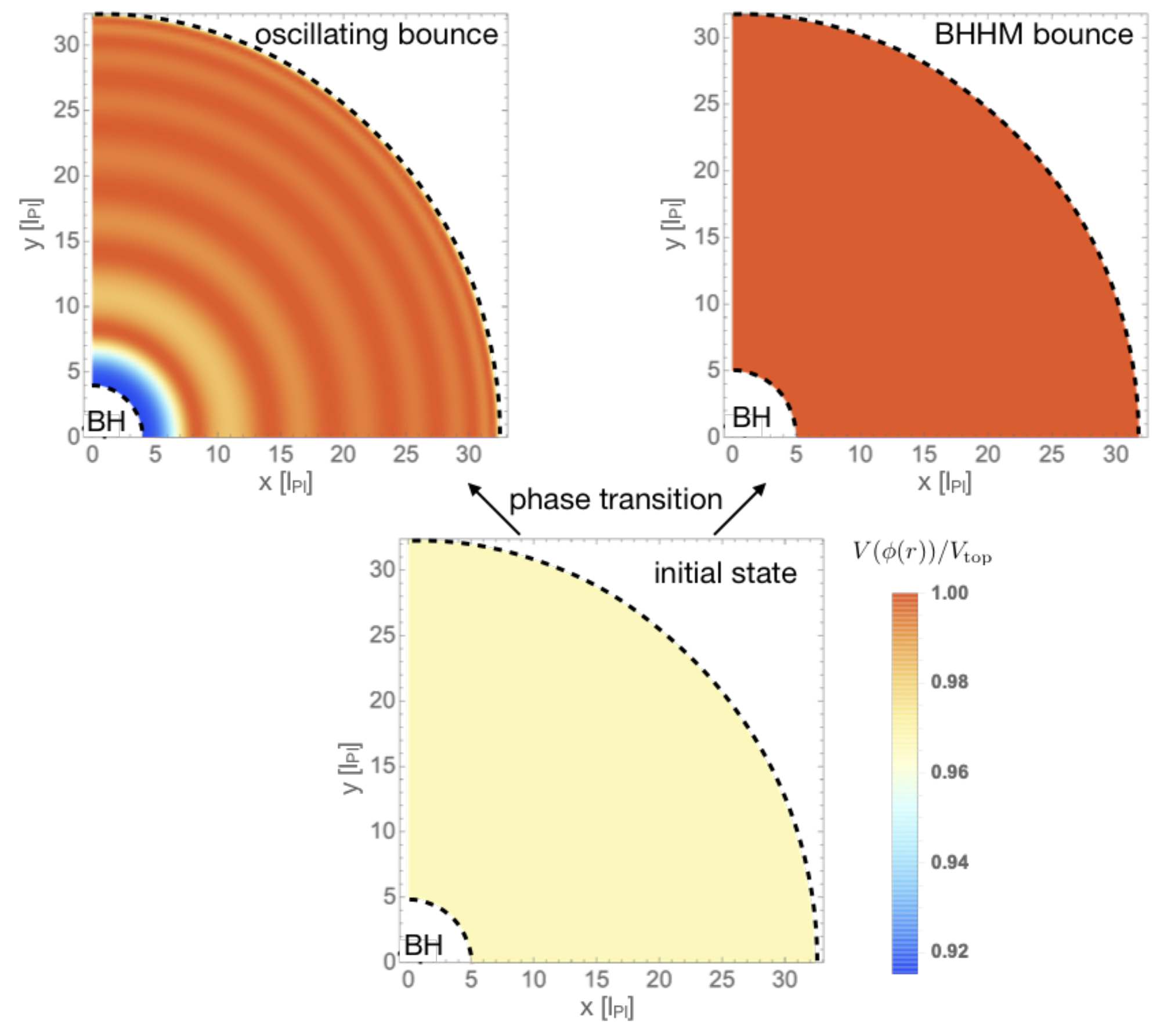}
\caption{The vacuum energy density profiles (color shows the value 
of $V(\phi (r))/V_{\text{top}}$) of the oscillating bounce ($k=7$), BHHM bounce, 
and the initial state. Dashed lines denote the BH and cosmological horizons. 
We set $\beta = 250$ and $\mu_- (G^3 V_{\rm top})^{1/2} = 0.02$.
}
\label{transition}
\end{figure}

The most important similarity between the oscillating bounce and HM 
bounce is the up-tunneling of matter fields since the oscillating bounce 
with a large $k$ has a very thick bubble wall dominated by the energy 
density at the top of the barrier (see FIG.\ \ref{transition}).
The main difference is that the pure HM instanton represents a 
jump in the global geometry due to the transition to the top of the potential,
which in turn results in a different cosmological horizon area. 
Once there is a BH however, it is now possible for the HM
bounce to have the same cosmological horizon geometry. In this section, 
we show that the transition rate of oscillating bounces is in good agreement 
with that of this type of HM bounce around a BH. In Sec.\ \ref{sec:define_BHHM} we 
discuss the HM bounce around a BH with fixed horizon geometry
(hereafter referred to as the fixed black 
hole Hawking-Moss (BHHM) bounce) and then we compare the oscillating 
bounce and BHHM bounce in Sec.\ \ref{sec:comparison}.

\subsection{HM bounce in the presence of a BH}
\label{sec:define_BHHM}

The HM bounce is a thermal transition of homogeneous dS space. 
Since a thermal state is an ensemble state of energy, up-tunneling is 
possible and its transition rate is suppressed by the Boltzmann factor. 
The HM bounce is very similar to this thermal transition and its decay rate 
can be also approximated by the Boltzmann factor
\begin{equation}
\Gamma \sim e^{-3/(8G^2V_{\rm min})+3/(8G^2V_{\rm top})} 
\simeq e^{-\Delta M_{\rm v}/T_{\rm ds}},
\label{BHHM_rate}
\end{equation}
where $V_{\rm min}$ and $V_{\rm top}$ denote the vacuum energy 
densities at a local minimum and top of potential, respectively, 
$\Delta M_{\rm v} \approx (4\pi/3)(V_{\rm top}-V_{\rm min})/H^3$
is notionally the mass-energy inside the cosmological horizon, 
and $T_{\rm ds} \approx H/2\pi$ is the mean de Sitter temperature
(assuming $V_{\rm min} \simeq V_{\rm top}$). 

Eq.\ (\ref{BHHM_rate}) means that the HM transition is suppressed by the 
increment of the total mass of vacuum energy, and this feature is very 
important when extending it to up-tunneling with a BH. 
This increment consists of both the BH and vacuum energy,
and so the up-tunneling rate 
should be suppressed by the increment of the total mass of both the 
BH and vacuum energy. If so, the up-tunneling rate may be 
enhanced when the decrement of the BH mass almost cancels out 
the increment of the vacuum energy. Recall, the fixed BHHM 
transition has an up-tunneling of vacuum energy while the total energy 
inside the cosmological horizon is conserved, i.e.\ the 
cosmological radii before and after the BHHM transition match:
\begin{align}
&1-\frac{2GM_+}{r_c} -\frac{8 \pi G V_{\rm min}}{3} r_c^2 = 0,\\
&1-\frac{2GM_-}{r_c} -\frac{8 \pi G V_{\rm top}}{3} r_c^2 = 0.
\end{align}
Then one can obtain the following relations
\begin{align}
&M_+ + \frac{4\pi}{3} r_c^3 V_{\rm min} = M_- + \frac{4\pi}{3} r_c^3 V_{\rm top},\\
&\Delta M_{\rm bh} \equiv M_- -M_+ = \frac{4 \pi}{3} r_c^3 
(V_{\rm top} -V_{\rm min}) \equiv - \Delta M_{\rm v},\label{conservation1}
\end{align}
where $r_c$ is the cosmological radius. Hence the mass of a remnant BH 
is uniquely determined given the values of $V_{\rm top}$, $V_{\rm min}$, 
and $M_+$. One can also read that the total mass is conserved from 
(\ref{conservation1}).

The decay rate of the BHHM bounce is obtained from the difference of 
the Bekenstein-Hawking entropy between the initial and final SdS 
vacua \cite{Gregory:2013hja,Oshita:2016oqn}. The radii of the BH 
and cosmological horizons before and after the transition are given by
\begin{align}
r_{h\pm} = \frac{2}{\sqrt{3} H_{\pm}} \cos{\left( \frac{\pi 
+ \cos^{-1} (3\sqrt{3} GM_{\pm} H_{\pm})}{3} \right)}, \label{ana_hori1} \\
r_{c\pm} =\frac{2}{\sqrt{3} H_{\pm}} \cos{\left( \frac{\pi 
- \cos^{-1} (3\sqrt{3} GM_{\pm}H_{\pm})}{3} \right)}\label{ana_hori2},
\end{align}
respectively, where the suffix of $+$ ($-$) denotes the quantities before 
(after) the BHHM transition and $r_{c +} = r_{c -}$ is satisfied. 
When $GM_{\pm} H_{\pm} \ll 1$, one can expand (\ref{ana_hori1}) as
\begin{align}
r_{h\pm} &\simeq 2GM_{\pm} + 8 G M_{\pm} (GM_{\pm} H_{\pm})^2,
\end{align}
and the difference between the BH horizon area before and after the 
BHHM transition, $\Delta A$, is approximately given by
\begin{equation}
\Delta A = 4 \pi (r_{h-}^2 - r_{h+}^2) \simeq 16 \pi G^2( M_-^2 -M_+^2),
\end{equation}
and the change of the Bekenstein-Hawking entropy is
\begin{equation}
\Delta S = \frac{\Delta A}{4 G} \simeq 4 \pi G (M_-^2 -M_+^2) 
= \frac{\Delta M_{\rm bh}}{T_{\rm av}}
= -\frac{\Delta M_{\rm v}}{T_{\rm av}},
\end{equation}
where $T_{\rm av} \equiv 1/(8 \pi G M_{\rm av})$, $M_{\rm av} \equiv (M_++M_-)/2$, 
and we used (\ref{conservation1}) in the last equality. Finally, we obtain the 
decay rate of the BHHM bounce as
\begin{equation}
\Gamma \sim e^{\Delta S} \simeq e^{\Delta M_{\rm bh}/T_{\rm av}} 
= e^{-\Delta M_{\rm v}/T_{\rm av}} \ \ \text{for} \ \ GM_{\pm} H_{\pm} \ll 1.
\end{equation}
Note that the BHHM transition consumes some of the mass of seed BH, 
therefore there is a threshold of seed mass below which there is no BHHM
instanton. This is precisely analogous to the critical seed mass in CDL
black hole seeded decay \cite{Gregory:2013hja}, which is the lowest
possible black hole seed that has a static decay instanton.

The lowest seed mass (hereafter referred to as the critical mass), 
below which the seed BH cannot catalyze the HM transition, can be 
obtained as follows. 
Expanding (\ref{ana_hori2}) with respect to 
$GM_{\pm} H_{\pm}$, one obtains
\begin{equation}
r_{\rm c \pm} \simeq \frac{1}{H_{\pm}} ( 1 - GM_{\pm}H_{\pm}),
\end{equation}
and requiring the energy conservation, i.e., $r_{\rm c +} = r_{\rm c -}$, 
one can obtain the following relation
\begin{equation}
\frac{1}{H_+}\left( 1 -GM_+H_+ +{\cal O} ((GM_+H_+)^2)\right) 
= \frac{1}{H_-}\left( 1 -GM_-H_- +{\cal O} ((GM_-H_-)^2)\right).
\end{equation}
Then the positivity of the remnant mass $(M_- \geq 0)$ gives the critical 
mass of $M_+$
\begin{equation}
M_+ \geq \frac{1}{GH_+} - \frac{1}{GH_-} \equiv M_{\rm c} \ \ \text{for} \ \ 
G M_{\pm} H_{\pm} \ll 1.
\label{critical_mass}
\end{equation}
At the critical mass $M_+ = M_{\rm c}$, the exponent of the decay rate is 
simply given by the Bekenstein-Hawking entropy of the seed BH 
$\Gamma \sim e^{-A/4G}$, where $A$ is the horizon area of the seed BH. 
In the next subsection, we come back to the oscillating bounce and we 
show some supporting evidences that the oscillating bounce with many 
oscillations corresponds to the BHHM bounce.

\subsection{Comparison between the oscillating bounce and BHHM bounce}
\label{sec:comparison}

\begin{figure}[h]
\centering
\includegraphics[width=0.8 \textwidth]{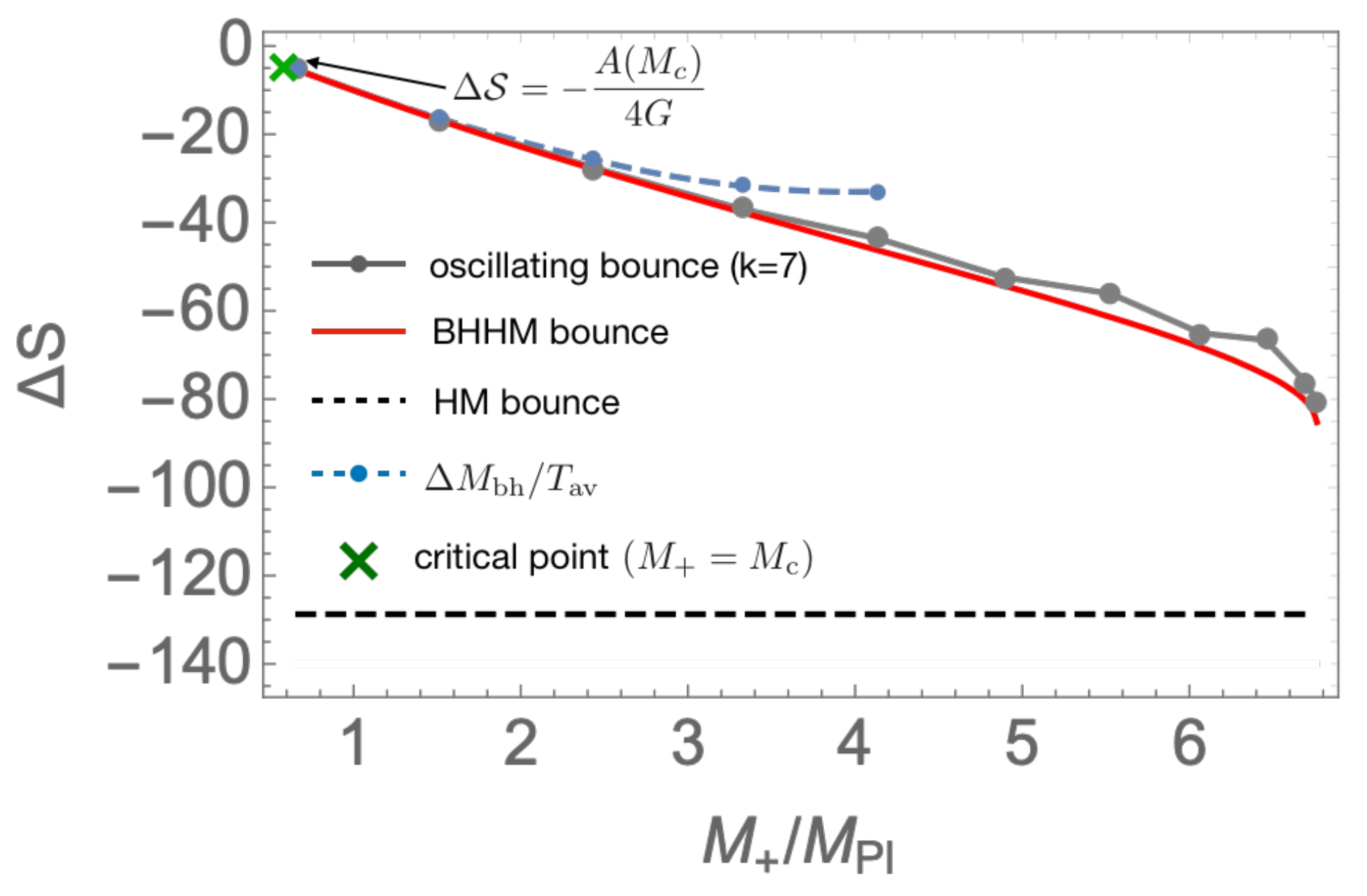}
\caption{Plot of the entropy change for the oscillating bounce with 
$k=7$ (grey), BHHM bounce (red) and HM bounce (black dashed), 
and $\Delta M_{\rm bh}/T_{\rm av}$ (blue). The critical point 
($M_+ =M_c \simeq 0.59 M_{\rm Pl}$) is marked with the green 
cross and the Nariai limit is at $M_+ \simeq 6.7 M_{\text{Pl}}$ in our setup. We set $\beta = 250$.
}
\label{entropy}
\end{figure}

Let us calculate the vacuum decay rate of the oscillating bounce and 
compare it with that of the fixed BHHM bounce. Since the oscillating bounce 
is a static solution, its on-shell Euclidean action is minus 
the Bekenstein-Hawking entropy. Therefore, the exponent of 
the transition rate is given by the entropy change before and after the 
phase transition $\Delta {\cal S}$. We numerically calculate $\Delta {\cal S}$ 
for the oscillating bounce of $k=7$. We find out that the vacuum decay 
rate for the oscillating bounce is in good agreement with that of the 
BHHM bounce (FIG.\ \ref{entropy}), and in the low mass limit 
($GM_+H_+ \ll 1$) it is consistent with the Boltzmann factor 
$e^{\Delta M_{\rm bh}/T_{\rm av}}$. It is also found out that the decay 
rate of the HM bounce with a BH is highest at the critical mass, for 
which $\Delta {\cal S} = -A (M_{\rm c})/4G$. Therefore, one can conclude 
that a smaller seed BH, but larger than the critical mass given in 
(\ref{critical_mass}), more strongly enhances the HM transition rate, and 
oscillating bounces with many oscillations is almost identical to the BHHM 
bounce from the point of view of the vacuum energy density profile 
(FIG.\ \ref{transition}) and of the decay rate (see a red and gray 
lines in FIG.\ \ref{entropy}).

\section{Conclusion}
\label{sec_conclusion}

We investigated the oscillating bounce solutions in the SdS background by 
using the static patch. We first considered the linearlized equation of motion 
of the scalar field around the top of the potential and demonstrated that the 
oscillating bounces in the dS background in both the global and static patches 
have the same eigenvalue restrictions on $\beta$. This is highly non-trivial 
since the Wick rotation in the global patch essentially differs from that in the 
static one. We also analytically obtained the oscillating bounce in the Nariai limit. 
To extend the bounce solutions to the general SdS case and to take into account 
the backreaction on spacetime and the non-linear terms in the equation of motion 
of $\phi$, we numerically solved the Einstein equation and the equation of motion
in the spherical and static case. Then we found out that the minimum value of 
$\beta$ allowing the existence of oscillating bounce with $k$ oscillations at the 
top of barrier, $\beta_k$, is restricted in the range $3k (k+1) \lesssim \beta_k 
\lesssim 2k (2k+3)$, where the lower and upper bounds on $\beta_k$ are 
associated with the Nariai and dS limits, respectively. We also checked the 
analytic bounce solutions in the dS and Nariai limit are consistent with the 
numerical solutions.

Since the oscillating bounce with many oscillations has its thick wall, it is 
natural to expect that such an oscillating bounce may be regarded as an 
intermediate bounce between the BHHM and static monotone ($k=1$) 
bounces around a BH. To check this, we estimated the action for the 
BHHM bounce and found out that it is consistent with the action of an 
oscillating bounce with many oscillations (comparison with $k=7$ is 
shown in FIG.\ \ref{entropy}). Remarkably, the BHHM transition rate is 
much higher than the HM transition rate without BH (black dashed line 
in FIG.\ \ref{entropy}). Note that the BH mass should be smaller than the 
Nariai mass and larger than the critical mass given in (\ref{critical_mass}). 
Therefore, we can conclude that a BH in the range of $1/G (H_+^{-1} - 
H_-^{-1}) < M_+ < (\sqrt{27} GH_-)^{-1}$ would catalyze the HM transition, 
and this effect is significant for small BHs close to the critical limit $M_+ \gtrsim M_c$.

The fact that there is a critical mass, below which the fixed BHHM instantons
do not exist is reminiscent of the situation with BH bounces 
\cite{Gregory:2013hja}, where there is critical mass below which static 
BH bounce solutions do not exist. In that case, the mass gap was completed
by having CDL-type instantons with no remnant mass. 
We plan to explore a wider class of instantons
beyond the fixed BHHM instantons considered here in the context of 
the oscillating bounce solutions.

\begin{acknowledgments}
This work was supported in part by the Leverhulme Trust (RG/IGM), 
STFC (Consolidated Grant ST/P000371/1) (RG/IGM), JSPS Overseas 
Research Fellowships (NO) and by the Perimeter Institute for 
Theoretical Physics.
Research at Perimeter Institute is supported in part by the Government of Canada through the Department of Innovation, Science and Economic Development Canada and by the Province of Ontario through the Ministry of Colleges and Universities.

\end{acknowledgments}

\providecommand{\href}[2]{#2}


\begin{thebibliography}{99}

\bibitem{coleman1977}
S.~Coleman, 
{\it {Fate of the false vacuum: Semiclassical theory}},  
Phys.Rev. {\bf D15} (1977) 2929--2936.

\bibitem{callan1977}
C. G. Callan and S.~Coleman,
{\it {Fate of the false vacuum II: First quantum corrections}},  
Phys.Rev. {\bf D16} (1977) 1762--1768.

\bibitem{CDL}
S.~Coleman and F.~De~Luccia, 
{\it {Gravitational effects on and of vacuum decay}},  
Phys.Rev. {\bf D21} (1980) 3305--3315.

\bibitem{Gregory:2013hja}
R.~Gregory, I.~G.~Moss and B.~Withers,
{\it {Black holes as bubble nucleation sites}},  
JHEP {\bf 1403} (2014) 081,
[\href{http://xxx.lanl.gov/abs/1401.0017}{{\tt arXiv:1401.0017 [hep-th]}}].

\bibitem{Burda:2015isa} 
P.~Burda, R.~Gregory and I.~Moss,
{\it Gravity and the stability of the Higgs vacuum},
Phys.\ Rev.\ Lett.\  {\bf 115}, 071303 (2015)
[\href{http://xxx.lanl.gov/abs/1501.04937}{{\tt arXiv:1501.024937 [hep-th]}}].

\bibitem{Burda:2015yfa}
P.~Burda, R.~Gregory and I.~Moss,
{\it Vacuum metastability with black holes},
JHEP {\bf 1508}, 114 (2015)
[\href{http://xxx.lanl.gov/abs/1503.07331}{{\tt arXiv:1503.07331 [hep-th]}}].

\bibitem{Burda:2016mou} 
P.~Burda, R.~Gregory and I.~Moss,
{\it The fate of the Higgs vacuum},
JHEP {\bf 1606}, 025 (2016)
[\href{http://xxx.lanl.gov/abs/1601.02152}{{\tt arXiv:1601.02152 [hep-th]}}].

\bibitem{Mukaida:2017bgd}
K.~Mukaida and M.~Yamada,
{\it False Vacuum Decay Catalyzed by Black Holes},
Phys.\ Rev.\ D {\bf 96} (2017) no.10,  103514
[\href{http://xxx.lanl.gov/abs/1706.04523}{{\tt arXiv:1706.04523 [hep-th]}}].

\bibitem{Oshita:2019jan} 
N.~Oshita, K.~Ueda and M.~Yamaguchi,
{\it Vacuum decays around spinning black holes},
JHEP {\bf 2001}, 015 (2020)
[\href{http://xxx.lanl.gov/abs/1909.01378}{{\tt arXiv:1909.01378 [hep-th]}}].

\bibitem{Koga:2019mee}
I.~Koga, S.~Kuroyanagi and Y.~Ookouchi,
{\it Instability of Higgs Vacuum via String Cloud}
Phys.\ Lett.\ B {\bf 800} (2020) 135093
[\href{http://xxx.lanl.gov/abs/1910.02435}{{\tt arXiv:1910.02435 [hep-th]}}].

\bibitem{Oshita:2018ptr}
N.~Oshita, M.~Yamada and M.~Yamaguchi,
{\it Compact objects as the catalysts for vacuum decays},
Phys.\ Lett.\ B {\bf 791} (2019) 149
[\href{http://xxx.lanl.gov/abs/1808.01382}{{\tt arXiv:1808.01382 [gr-qc]}}].

\bibitem{Hiscock:1987hn}
W.~A. Hiscock,
{\it {Can black holes nucleate vacuum phase transitions?}},  
Phys.Rev. {\bf D35} (1987) 1161--1170.

\bibitem{Berezin:1987ea}
V.~Berezin, V.~Kuzmin, and I.~Tkachev, 
{\it {O(3) invariant tunneling in general relativity}},  
Phys.Lett. {\bf B207} (1988) 397.

\bibitem{Berezin:1990qs} 
V.~A.~Berezin, V.~A.~Kuzmin and I.~I.~Tkachev,
{\it Black holes initiate false vacuum decay},
{\em Phys.\ Rev.\ D} {\bf 43}, 3112 (1991).

\bibitem{Hawking:1981fz} 
S.~W.~Hawking and I.~G.~Moss,
{\it Supercooled Phase Transitions in the Very Early Universe},
Phys.\ Lett.\  {\bf 110B}, 35 (1982)
[Adv.\ Ser.\ Astrophys.\ Cosmol.\  {\bf 3}, 154 (1987)].

\bibitem{Hartle:1983ai}
J.~B.~Hartle and S.~W.~Hawking,
{\it Wave Function of the Universe}
Phys.\ Rev.\ D {\bf 28} (1983) 2960
[Adv.\ Ser.\ Astrophys.\ Cosmol.\  {\bf 3} (1987) 174]

\bibitem{Starobinsky:1986fx}
A.~A.~Starobinsky,
{\it Stochastic De Sitter (inflationary) Stage In The Early Universe}
Lect.\ Notes Phys.\  {\bf 246} (1986) 107.

\bibitem{Brown:2007sd}
A.~R. Brown and E.~J. Weinberg, 
{\it {Thermal derivation of the Coleman-De Luccia tunneling prescription}},
Phys.Rev. {\bf D76} (2007) 064003,
[\href{http://xxx.lanl.gov/abs/0706.1573}{{\tt arXiv:0706.1573 [hep-th]}}].

\bibitem{Hackworth:2004xb} 
J.~C.~Hackworth and E.~J.~Weinberg,
{\it Oscillating bounce solutions and vacuum tunneling in de Sitter spacetime},
Phys.\ Rev.\ D {\bf 71}, 044014 (2005)
[\href{http://xxx.lanl.gov/abs/hep-th/0410142}{{\tt  hep-th/0410142}}].
  
\bibitem{Weinberg:2005af} 
E.~J.~Weinberg,
{\it New bounce solutions and vacuum tunneling in de Sitter spacetime},
AIP Conf.\ Proc.\  {\bf 805}, no. 1, 259 (2005)
[\href{http://xxx.lanl.gov/abs/hep-th/0512332}{{\tt  hep-th/0512332}}].

\bibitem{Balek:2004sd} 
V.~Balek and M.~Demetrian,
{\it Euclidean action for vacuum decay in a de Sitter universe},
Phys.\ Rev.\ D {\bf 71}, 023512 (2005)
[\href{http://xxx.lanl.gov/abs/gr-qc/0409001}{{\tt  gr-qc/0409001}}].

\bibitem{Battarra:2013rba} 
L.~Battarra, G.~Lavrelashvili and J.~L.~Lehners,
{\it Zoology of instanton solutions in flat potential barriers},
Phys.\ Rev.\ D {\bf 88}, 104012 (2013)
[\href{http://xxx.lanl.gov/abs/1307.7954}{{\tt arXiv:1307.7954 [hep-th]}}].

\bibitem{Kanno:2011vm} 
S.~Kanno and J.~Soda,
{\it Exact Coleman-de Luccia Instantons},
Int.\ J.\ Mod.\ Phys.\ D {\bf 21}, 1250040 (2012)
[\href{http://xxx.lanl.gov/abs/1111.0720}{{\tt arXiv:1111.0720 [hep-th]}}].

\bibitem{Demetrian:2005ag} 
M.~Demetrian,
{\it False vacuum decay with gravity in a critical case},
Int.\ J.\ Theor.\ Phys.\  {\bf 46}, 652 (2007)
[\href{http://xxx.lanl.gov/abs/gr-qc/0504133}{{\tt  gr-qc/0504133}}].

\bibitem{Stopyra:2018cjy}
S.~Stopyra,
{\it Standard Model Vacuum Decay with Gravity}, 
Ph. D. thesis.

\bibitem{Rajantie:2016hkj}
A.~Rajantie and S.~Stopyra,
{\it Standard Model vacuum decay with gravity},
Phys.\ Rev.\ D {\bf 95} (2017) no.2,  025008
[\href{http://xxx.lanl.gov/abs/1606.00849}{{\tt arXiv:1606.00849 [hep-th]}}].

\bibitem{Joti:2017fwe}
A.~Joti, A.~Katsis, D.~Loupas, A.~Salvio, A.~Strumia, N.~Tetradis and A.~Urbano,
{\it (Higgs) vacuum decay during inflation},
JHEP {\bf 1707} (2017) 058
[\href{http://xxx.lanl.gov/abs/1706.00792}{{\tt arXiv:1706.00792 [hep-th]}}].

\bibitem{Lavrelashvili:2006cv} 
G.~Lavrelashvili,
{\it The Number of negative modes of the oscillating bounces},
Phys.\ Rev.\ D {\bf 73}, 083513 (2006) 
[\href{http://xxx.lanl.gov/abs/gr-qc/0602039}{{\tt  gr-qc/0602039}}].

\bibitem{Lee:2012qv} 
B.~H.~Lee, W.~Lee and D.~h.~Yeom,
{\it Oscillating instantons as homogeneous tunneling channels},
Int.\ J.\ Mod.\ Phys.\ A {\bf 28}, 1350082 (2013)
[\href{http://xxx.lanl.gov/abs/1206.7040}{{\tt arXiv:1206.7040 [hep-th]}}].

\bibitem{Lee:2014ula}
B.~H.~Lee, W.~Lee, D.~Ro and D.~h.~Yeom,
{\it Oscillating Fubini instantons in curved space},
Phys.\ Rev.\ D {\bf 91} (2015) no.12,  124044
[\href{http://xxx.lanl.gov/abs/1409.3935}{{\tt arXiv:1409.3935 [hep-th]}}].

\bibitem{Oshita:2016oqn} 
N.~Oshita and J.~Yokoyama,
{\it Entropic interpretation of the Hawking--Moss bounce},
PTEP {\bf 2016}, no. 5, 053E02 (2016)
[\href{http://xxx.lanl.gov/abs/1603.06671}{{\tt arXiv:1603.06671 [hep-th]}}].

\bibitem{Moss:1984zf}
I.~G.~Moss,
{\it Black Hole Bubbles},
Phys.\ Rev.\ D {\bf 32} (1985) 1333.

\bibitem{Flachi:2011sx}
A.~Flachi and T.~Tanaka,
{\it Chiral Phase Transitions around Black Holes},
Phys.\ Rev.\ D {\bf 84} (2011) 061503
[\href{http://xxx.lanl.gov/abs/1106.3991}{{\tt arXiv:1106.3991 [hep-th]}}].

\bibitem{Oshita:2016btk}
N.~Oshita and J.~Yokoyama,
Phys.\ Lett.\ B {\bf 785} (2018) 197
[\href{http://xxx.lanl.gov/abs/1601.03929}{{\tt arXiv:1601.03929 [gr-qc]}}].

\bibitem{Nariai:1}
H.~Nariai, 
{\it On a new cosmological solution of Einstein's field equations of gravitation}, 
Sci.\ Rep.\ Tohoku Univ.\ Ser.\ 1, (1951). 
Gen.~Rel.~Grav., 31, 963-971 (1999).

\bibitem{Nariai:2}
H.~Nariai, 
{\it On some static solutions of Einstein's gravitational field equations in a 
spherically symmetric case}, 
Sci.\ Rep.\ Tohoku Univ.\ Eighth Ser., (1950). 
Gen.~Rel.~Grav., 31, 951-961 (1999).

\bibitem{Gomberoff:2005je} 
A.~Gomberoff, M.~Henneaux and C.~Teitelboim,
{\it Decay of the cosmological constant: Equivalence of quantum tunneling 
and thermal activation in two spacetime dimensions},
Phys.\ Rev.\ D {\bf 71}, 063509 (2005)
[\href{http://xxx.lanl.gov/abs/hep-th/0501152}{{\tt  hep-th/0501152}}].

\bibitem{Masoumi:2012yy} 
A.~Masoumi and E.~J.~Weinberg,
{\it Bounces with O(3) x O(2) symmetry},
Phys.\ Rev.\ D {\bf 86}, 104029 (2012)
[\href{http://xxx.lanl.gov/abs/1207.3717}{{\tt arXiv:1207.3717 [hep-th]}}].

\bibitem{Ai:2018rnh} 
W.~Y.~Ai,
{\it Correspondence between Thermal and Quantum Vacuum Transitions 
around Horizons},
JHEP {\bf 1903}, 164 (2019)
[\href{http://xxx.lanl.gov/abs/1812.06962}{{\tt arXiv:1812.06962 [hep-th]}}].

\bibitem{Battarra:2012vu}
L.~Battarra, G.~Lavrelashvili and J.~L.~Lehners,
{\it Negative Modes of Oscillating Instantons},
Phys. Rev. D \textbf{86} (2012), 124001
[\href{http://xxx.lanl.gov/abs/1208.2182}{{\tt arXiv:1208.2182 [hep-th]}}].

\end{thebibliography}
\end{document}